\documentclass[sn-mathphys,Numbered]{sn-jnl}

\usepackage{graphicx}%
\usepackage{multirow}%
\usepackage{amsmath,amssymb,amsfonts}%
\usepackage{amsthm}%
\usepackage{mathrsfs}%
\usepackage[title]{appendix}%
\usepackage{xcolor}%
\usepackage{textcomp}%
\usepackage{manyfoot}%
\usepackage{booktabs}%
\usepackage{algorithm}%
\usepackage{algorithmicx}%
\usepackage{algpseudocode}%
\usepackage{listings}%
\usepackage{lineno}

\theoremstyle{thmstyleone}%
\theoremstyle{thmstyletwo}%

\theoremstyle{thmstylethree}%

\begin{document}

\title[Article Title]{Integrated photonics modular arithmetic processor} 


\author[1]{\fnm{Yuepeng} \sur{Wu}}\email{ketchup98@bupt.edu.cn}

\author[1]{\fnm{Hongxiang} \sur{Guo}}\email{hxguo@bupt.edu.cn}
\author[1]{\fnm{Bowen} \sur{Zhang}}\email{bwz1997@bupt.edu.cn}
\author[1]{\fnm{Jifang} \sur{Qiu}}\email{jifangqiu@bupt.edu.cn}
\author[1]{\fnm{Zhisheng} \sur{Yang}}\email{zhisheng.yang@bupt.edu.cn}
\author*[1]{\fnm{Jian} \sur{Wu}}\email{jianwu@bupt.edu.cn}

\affil*[1]{\orgdiv{School of Electronic Engineering}, \orgname{Beijing University of Posts and Telecommunications}, \orgaddress{\city{Beijing}, \postcode{100876}, \country{China}}}




\abstract{Integrated photonics computing has emerged as a promising approach to overcome the limitations of electronic processors in the post-Moore era, capitalizing on the superiority of photonic systems.
However, present integrated photonics computing systems face challenges in achieving high-precision calculations, consequently limiting their potential applications, and their heavy reliance on analog-to-digital (AD) and digital-to-analog (DA) conversion interfaces undermines their performance.
Here we propose an innovative photonic computing architecture featuring scalable calculation precision and a novel photonic conversion interface.
By leveraging Residue Number System (RNS) theory, the high-precision calculation is decomposed into multiple low-precision modular arithmetic operations executed through optical phase manipulation. Those operations directly interact with the digital system via our proposed optical digital-to-phase converter (ODPC) and phase-to-digital converter (OPDC).
Through experimental demonstrations, we showcase a calculation precision of 9 bits and verify the feasibility of the ODPC/OPDC photonic interface.
This approach paves the path towards liberating photonic computing from the constraints imposed by limited precision and AD/DA converters.
}




\maketitle

\section{Introduction}\label{sec1}

\begin{figure}[t!]
	\centering\includegraphics[width=\linewidth]{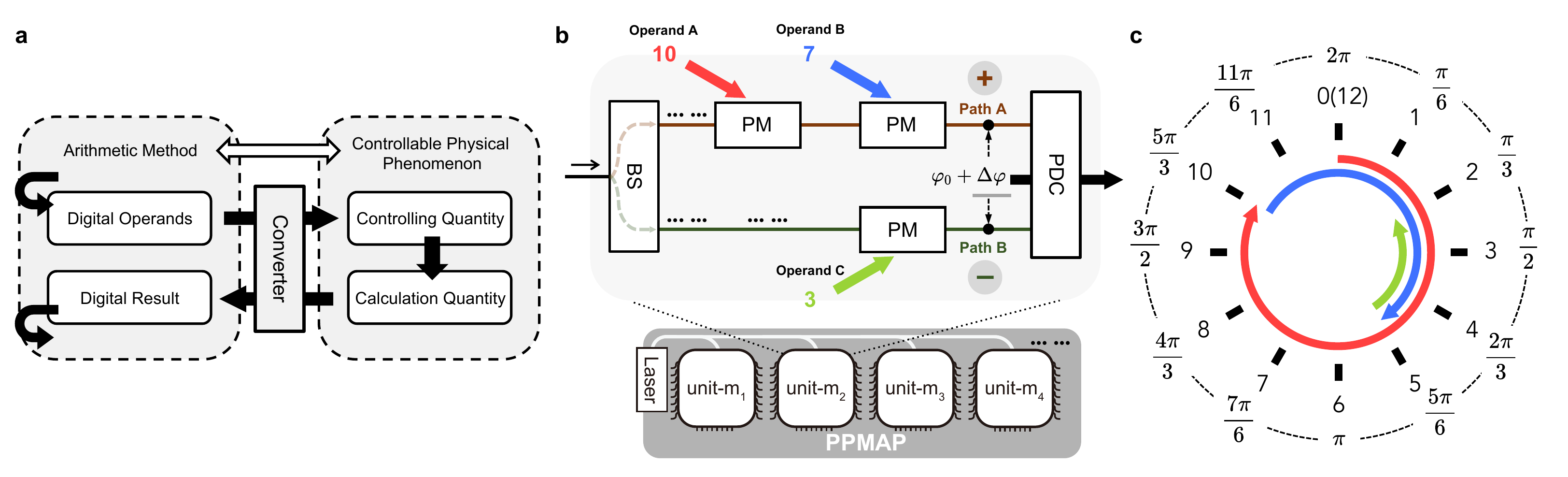}
	\caption{
	(a) General structure of emerging photonic computing. 
    For a controllable physical phenomenon, once a ``controlling quantity'' is specified, the ``calculation quantity'' evolves rapidly based on a specific physical relationship. By exploiting this relationship and its inherent correspondence between a specific arithmetic method, efficient calculations can be achieved through the loading and extraction of digital information via converters.
	(b) Schematic diagram of the PPMAP architecture. 
	(c) The mapping between operands and phase when take modulo-12 as an example, along with a demonstration of basic modular arithmetic performed based on this mapping. \label{fig1}
	}
\end{figure}

Since the breakdown of Dennard scaling of integrated circuits around 2006, making individual computing units faster and more energy-efficient has become increasingly challenging
\cite{hennessy2018new}.
This has led to current processors being limited to clock frequencies of a few GHz and commonly experiencing low utilization due to restricted power dissipation \cite{dark_silicon}.
These challenges are the direct consequence of the fundamental physics problems that CMOS transistor technology faces, which results in computing performance being a major bottleneck for many potential applications \cite{thompson2022importance}. 
Therefore, many efforts have been made to improve the conventional transistor-based computing paradigm with an alternative platform.
One of the promising approaches is integrated photonic computing, which enables operation at unprecedented bandwidths \cite{2007microwave}. 
To fully harness the superiority of the photonic platform and construct a competitive computing system, the primary challenge is to find a natural and efficient combination between the arithmetic method and controllable physical effect in integrated photonics, just like the combination between Boolean algebra and transistors (as illustrated in Fig. \ref{fig1}a).

Currently, most published integrated photonics computing research utilizes intensity as the controllable physical quantity, i.e., ``calculation quantity'', and adopts arithmetic methods that mimic their electronic counterparts.
Some researches aimed at replacing transistor logic gates in the electric Boolean digital system with photonic ones \cite{yang2018nanowire, caballero2022photonic}. However, optical intensity signals suffer the loss from both the material absorption and the intrinsic property of non-reciprocal Boolean operation \cite{ying2020electronic}
, which 
makes it challenging to be cascaded to form a photonic arithmetic unit with sufficiently high bit-width. 
On the other hand,
more researches adopted the paradigm of neuromorphic computing \cite{sebastian2020memory, shafiee2016isaac, mehonic2022brain},
which aims to implement
matrix-vector multiplication \cite{photonicMAC, filipovich2022silicon, pai2023experimental},
convolution operation \cite{meng2023compact, zhou2023memory, xu2022high}, 
or neural network model inference \cite{nature2022onn, shen2017deep, ashtiani2022chip, fu2023photonic}
on integrated optical devices
with ultra-fast calculation speed. 
However, 
the analog intensity signal is susceptible to noise, which limits their calculation precision due to relatively poor signal-to-noise ratio (SNR) \cite{tait2022quantifying}.
The calculation precision of these published optical neuromorphic systems cannot meet the general requirements
\cite{xu202111, krishnamoorthi2018quantizing}, and there is no efficient method to improve their calculation precision as compared to digital computation systems \cite{garg2022dynamic}.
Consequently, its demonstrated workloads are severely restricted to tasks of simplistic functionality, undermining the potential of integrated photonics computing for broader application domains.
Additionally, to ensure compatibility with existing digital devices, an optical neuromorphic system operating at ultra-high bandwidth requires high-precision analog-to-digital converters (ADCs) and digital-to-analog converters (DACs) that can match its speed. 
This requirement poses significant energy consumption challenges \cite{adc_survey, morales2022analysis} while also imposing formidable limitations on the performance of the integrated photonics computing system.
These ``precision challenge'' and ``converter challenge'' have hindered the practical implementation of the intensity-based methods and have necessitated the pursuit of alternative calculation quantity and arithmetic methods for integrated photonics computing.

Phase is another controllable physical quantity in integrated photonics apart from the intensity and has not yet been fully explored in the field of photonic computing, which can be controlled more efficiently by integrated phase modulator (PM) \cite{dong2022high, kharel2021breaking} and exhibiting intrinsic periodicity.
That feature of the optical phase aligns naturally with the modular arithmetic in which numbers are folded by a specific modulus \cite{nicholson2012introduction}.
For modulus arithmetic, leveraging a non-traditional numerical representation format, i.e., residue number system (RNS), allows for the decomposition of high-precision computations into multiple parallel low-precision operations \cite{rrns2007residue}.
The correspondence between the modular arithmetic and the optical phase provides a foundation for leveraging the properties of modular arithmetic in the efficient extension of calculation precision in integrated photonic computing systems.
Here, we propose an integrated phase-based photonics modular arithmetic processor (PPMAP) architecture aimed at addressing the precision and converter challenges in photonics computing.
By taking advantage of the unique characteristics of phase, its components, PPMAP units, enable accurate execution of basic modular arithmetic operations and support multi-operand operations that are challenging to implement in electronic processors.
The precision of this architecture can be effectively enhanced using the RNS theory.
In a proof-of-concept experiment, the three-unit PPMAP achieved 9-bit calculation precision, which is unprecedented for intensity-based photonic computing architecture. 
Moreover, based on the multi-operand characteristics of PPMAP architecture and our previous work on phase-shifted optical quantizer\cite{liu2021experimental, tian2018chip}, we demonstrate the accurate loading and extraction of information into and from phase domain with the seamlessly integrated photonics conversion interfaces, which make advancement in elimination the constraints of AD/DA converters.
Further simulations indicate that PPMAP can achieve comparable reliability to electronic processors under reachable SNR conditions.

\section{Results}\label{sec2}
\subsection{PPMAP architecture}\label{subsec0}
The basic structure of the PPMAP unit is depicted in Fig.\ref{fig1}b. A beam splitter (BS) evenly splits the power of an incident monochromatic laser into two beams, which propagate through two separate paths. Each path is equipped with a set of cascaded phase modulators (PMs). Finally, the phase difference between the signals in the two paths is extracted by a phase-to-digital converter (PDC) linked at the end.
Since both beams originate from the same laser source, in the absence of a voltage bias on the modulator, they exhibit a stable static phase difference denoted as $\varphi_{0}$.
When a voltage is applied to one of the modulators, it introduces a dynamic phase difference denoted as $\Delta \varphi$.
In the case that simultaneously applies voltage to the PMs, the overall dynamic phase difference is the sum of their individual contributions.

To perform modular arithmetic operations under modulus $m$, we assign $m$ discrete phase points from the analog quantity $\Delta \varphi \in [0, 2\pi)$ to represent the $m$ elements in the set $ \{0,\ 1,\ 2,\ldots,m-1\} $. 
These points should include the phase of 0 as the identity element and be uniformly spaced with an interval of $2\pi/m$ to satisfy the closure property, which ensures that the result of the addition operation remains within the same set \cite{nicholson2012introduction}. 
Consequently, the integers in modulo $m$ are mapped to the discrete dynamic phase differences $\Delta \varphi_{\mathbb{Z}}=\{0, 2\pi/m, 4\pi/m, ..., 2\pi-2\pi/m\}$.
This mapping allows for modular addition or subtraction operations by loading operands into the corresponding phase modulators on the same or different paths, as shown in Fig.\ref{fig1}c, which illustrates an example of $m=12$. 
To perform a modular addition operation with operands ``10" and ``7", followed by subtraction with ``3", we regulate two PMs on the same path and one PM on the other path to induce phase changes of $5\pi/3$, $7\pi/6$, and $\pi/2$, respectively. 
The closure property allows us to derive the results once we identify the specific discrete phase points from $\Delta \varphi_{\mathbb{Z}}$. The PDC then determines the resulting phase $\Delta \varphi = 11\pi/6$, corresponding to the calculation result of "$\vert 10+7-3\vert_{12}=2$".
With that efficient implementation of modular addition, we can implement modular multiplication with the same structure through the method of index-sum multiplication that performs modular multiplication by simply a modular addition and table lookup operation \cite{rrns2021scalable}.

Additionally, by leveraging the theory of residue number system (RNS), which represents the integer operand, denoted as $X$, by its remainders of a set of pairwise coprime moduli $ \{ m_1,...,m_N \} $, i.e., the RNS format:

\begin{equation}
    X_{\rm RNS} = X \% \{ m_1,\dots,m_N \} = \{ | X|_{m_1},\dots, | X|_{m_N} \}
\end{equation}

, the original operation with high bit-width operands can be broken into multiple parallel modular arithmetic operations with much lower bit-width operands and different moduli, as long as the result is within the range $ M=\prod_{i=1}^{N}m_i $.
By orchestrating multiple PPMAP units with different moduli according to the RNS theory, as shown in Fig.\ref{fig1}b, we construct the PPMAP architecture with bit-width of $ \lfloor \log_{2} \prod_{i=1}^{N} m_{i} \rfloor $, thereby realizing calculation with scalable precision.

\begin{figure}[ht!]
	\centering\includegraphics[width=\linewidth]{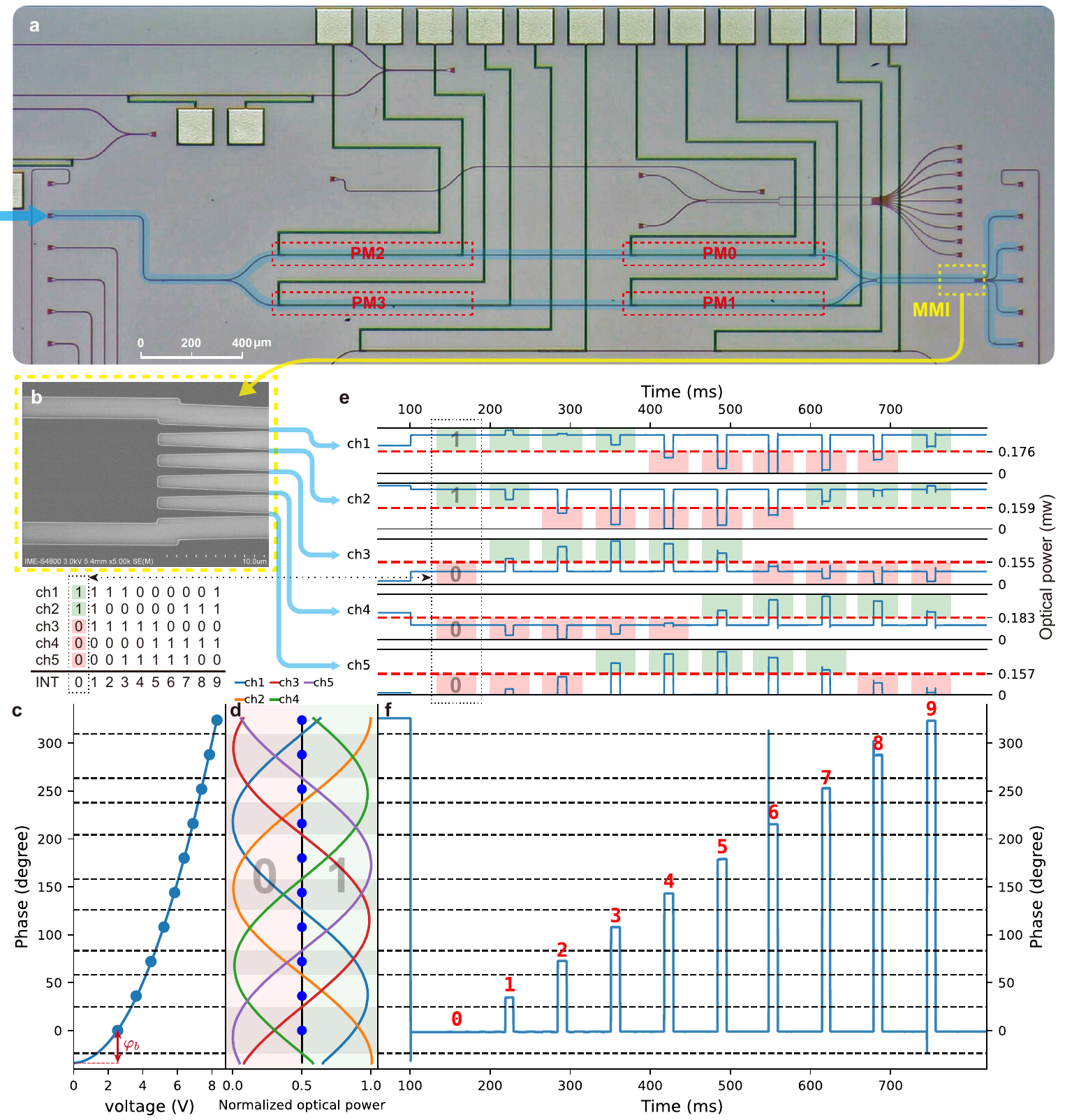}
	\caption
	{
	(a) Micrograph of the integrated photonic chip and 
	(b) MMI-structured OPDC, with five output channels ($K = 5$).
	(c) The relationship between the voltage applied to PM0, the magnitude of consequential phase change, and (d) the corresponding optical power of the output signals. 
	The zero phase reference point is chosen as $\varphi_b$, while the optical power signals from multiple channels were normalized relative to optimized thresholds, ensuring that all thresholds are aligned at 0.5 on the graph.
	(e) The multi-channel optical power signals along with their corresponding threshold decision results (red or green for 0 or 1, respectively) at each time slot, and (f) the phase signal reconstructed from the optical power signals. \label{fig2}
	}
\end{figure}

\subsection{High-precision integrated photonic calculation}\label{subsec1}

Fig.\ref{fig2}a illustrates the integrated photonics chip utilized for the principle verification experiments, which was fabricated on a commercially available 180-nm SOI platform via the multi-project wafer (MPW) process and corresponds to the fundamental structure depicted in Fig.\ref{fig1}b. The continuous wave (CW) laser is coupled into the integrated waveguide through a grating coupler and is evenly separated via a subsequent $1\times 2$ multimode interferometer (MMI).
Those two laser beams then respectively traverse two adjacent waveguides where four thermal PMs load operands into the produced phase difference.
The magnitude of the phase difference induced by a specific voltage value applied to the PM can be precisely controlled by the method described in Appendix\ref{secA1}.
Then, the two laser beams with a phase difference carrying the calculation result of those operands enter into an MMI-structured optical PDC (OPDC) to identify and quantify phase information.
Fig.\ref{fig2}b displays a micrograph of the OPDC with five output channels where the two input laser beams take place interference based on the self-imaging effect \cite{soldano1995optical}. 
That makes the optical power at each output channel capture different profiles of the phase information $\Delta \varphi$.
The relationship between the phase changes induced by varied control voltages applied to a modulator, PM0, (as depicted in Fig.\ref{fig2}c) and the optical power of the five channels is shown in Fig.\ref{fig2}d.
Performing threshold decision on the optical power signal of a channel using comparators divides the 360-degree phase range into multiple phase intervals.
This division occurs due to a change in the binary codewords (0s and 1s) generated by that channel.
By performing decision on the optical power of each channel, the specific phase interval to which the current phase belongs can be determined according to the obtained Gray code digital output \cite{tian2018chip}.
For accurately discriminating the information encoded in the phase, we introduce an additional static phase bias $\varphi_{b}$ and optimize it along with the decision thresholds 
(see Appendix\ref{secA2}).

Fig.\ref{fig2}c-f demonstrates the experimental process of the loading and extracting phase information for the case of $m=10$.
The voltages corresponding to different discrete phases in $\varphi_{b}+\Delta \varphi_{\mathbb{Z}}$ are applied to the modulator in varying time slots. 
By recovering the phase value from the multiple-channel optical power signals in Fig.\ref{fig2}e, it can be observed that the phase is controlled to be in the middle of the expected phase intervals, as shown in Fig.\ref{fig2}f.
Furthermore, based on the digital results of the threshold decision to the optical power of multiple channels, the specific discrete phase and its corresponding integer value for each time slot are discerned.
The methods of phase information loading and extraction described above are applicable to different moduli. 
This flexibility enables the instantiation the PPMAP units corresponding to different moduli using a single integrated photonic device.

\begin{figure}[t!]
	\centering\includegraphics[width=\linewidth]{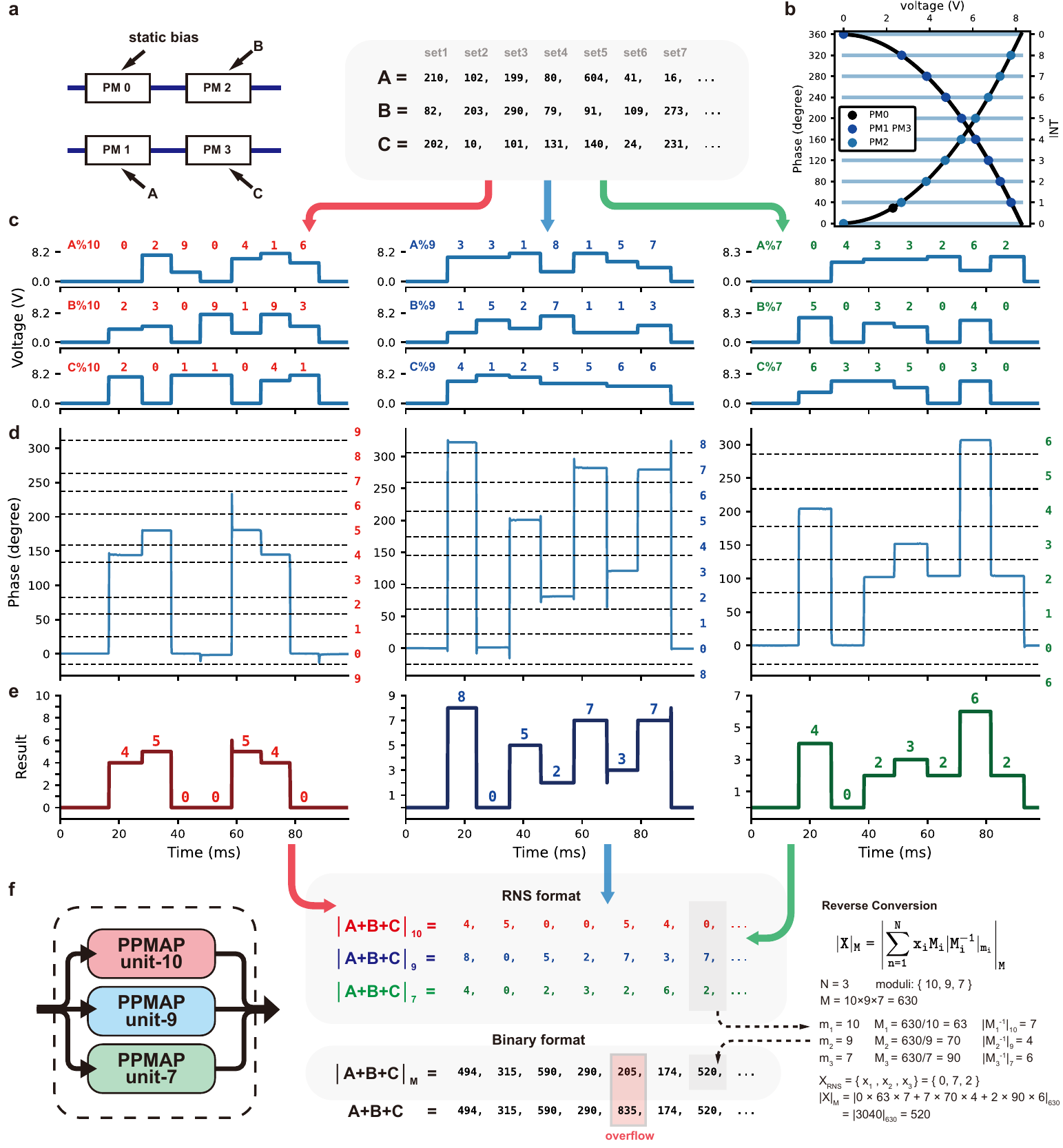}
	\caption
	{
	(a) The operand or function assigned to each modulator. 
	(b) The method of mapping operands to corresponding phases according to the voltage-phase relationship of each modulator, taking the case of $m$=9 as an example.
	(c-e) The parallel photonics-accelerated modular arithmetic operations in moduli $\{10, 9, 7\}$
	In different time slots, applying (c) voltage signals corresponding to RNS-formatted operands to the modulators to perform operations on different sample sets, resulting in (d) output phase signals and (e) quantized results obtained from the OPDC.
	(f) Schematic of the PPMAP architecture which contains three units corresponding to the moduli employed in the experiments.
	It is noted that there is an unconformity between the RNS result from the photonic processor and the binary result from the electric one in set5, which is due to integer overflow occurring in the calculation.
	} \label{fig3}
\end{figure}



We further evaluated the practicality of PPMAP architecture by experimentally implementing a 9-bit three-operand addition operation using moduli $\{10, 9, 7\}$. During the experiment, the same physical PPMAP unit depicted in Fig\ref{fig2} performed all the calculations involving different moduli, i.e., implementing three PPMAP units utilized one device via time division multiplexing.
Moreover, the optimization of $\varphi_{b}$ and $\vec{d}$ for each modulus was performed beforehand.
We selected seven sets of independent operand samples for this demonstration and performed $A+B+C$ operation on each set separately.
The bit width of the binary operands is $\left \lfloor \log_{2} \left( 10 \times 9 \times 7 \right) \right \rfloor=9$ bit.
After converting into RNS format, the original multi-operand addition operation transformed into three corresponding modular arithmetic operations with different moduli of $\{10, 9, 7\}$.

For each specific modulus, the operations on different sample sets perform consecutively at varying time slots.
For implementing the operation of a single set in the phase domain, we simultaneously apply the voltages corresponding to the converted operands to multiple PMs.
As shown in Fig.\ref{fig3}a,
we apply a constant voltage to PM0 to assign it the responsibility of providing the bias phase $\varphi_{b}$.
PM1, PM2, and PM3 are also assigned to accurately map operands A, B, and C to their respective phase change amounts.
The mapping method in the case of $m=9$ is illustrated in Fig.\ref{fig3}b as an example: 
since PM2 and PM0 locate on identical waveguides and $\varphi_{b}$ has been provided by PM0, we directly control the voltage of PM1 to induce a phase change of $ \Delta \varphi_{X} = X \cdot 2\pi/m $ for operand $X$. 
On the other hand,
since PM1 and PM3 are located on different waveguides than PM0, it is necessary to control the voltage to generate a phase change equal to $ 2\pi - \Delta \varphi_{X} $ when mapping X to phase $ \Delta \varphi_{X} $.
The applied voltage signals are shown in Fig.\ref{fig3}c, and their consequent calculation results in the phase domain are illustrated in Fig.\ref{fig3}d.
For a specific time slot, the phase values fall within the phase interval corresponding to the exact computed results, performing the desired calculations in the phase domain.
Fig.\ref{fig3}e displays the integer signal generated by the OPDC, which infers the precise extraction of the computation result in RNS format from the phase domain.

The RNS result can be mapped into the binary format using the reverse conversion, such as the Chinese Remainder Theorem (CRT) algorithm as demonstrated in Fig.\ref{fig3}.
Calculations performed in RNS format agree with binary ones as long as the final result does not exceed $M=630$.
It is worth mentioning that the relatively complex reverse conversion process is unessential. It can be avoided by carefully selecting the workload and employing advanced residue interaction methods \cite{rrns2007residue}. 
Moreover, recent research in computer architecture verifies that, including memory hierarchy, all computer parts can efficiently run within the RNS format \cite{rrns2018memory}.

\subsection{Optical Digital-to-Phase Converters}\label{subsec3}

To truly resolve the ``converter challenge'', in addition to the OPDC, dedicated electronic-photonic DAC modules that can convert digital operands into the analog phase domain are crucial.
Since a DAC essentially performs a weighted summation with fixed weight on a digital operand, denoted as $X$, and concurrently transfers the result into analog quantity $\Psi$: 

\begin{equation}
	X = \left(x_{L-1} x_{L-2} \ldots x_{0}\right)_{2} \rightarrow \Psi_{X} \left(\sum_{i=0}^{L-1} 2^{i} x_{i} \right)
\end{equation}

where $x_i=0 \text{ or } 1$ and $L$ represents the bit-width of $X$, we can leverage the just-demonstrated multi-operand feature of PPMAP architecture to implement the optical digital-to-phase converter (ODPC).

An ODPC comprises a collection of PMs positioned along the paths of the PPMAP unit. It receives the digital signal of one operand from the parallel bus.
In the basic design of the ODPC, it is equipped with $L$ PMs that are respectively controlled by each binary digit $ x_i $ in the $ X $.
Each PM in the ODPC is engineered to exhibit a specific phase response to the voltage amplitude of the digital signal.
In particular, when the phase response of the $i^{th}$ PM is designed as $ \Delta \varphi_i = 2\pi | 2^i |_m / m $, the OPDC enables the direct transformation of the $L$ synchronously applied digital signal into the resulting summation in the phase domain:

\begin{equation} \label{equation_3}
    \Delta \varphi_{X}=\sum_{i=0}^{L-1} x_{i} \Delta \varphi_i =\sum_{i=0}^{L-1} x_{i} | 2^i |_{m} \frac{2 \pi}{m}=\frac{2 \pi}{m} \sum_{i=0}^{L-1} | 2^{i} x_{i}|_{m}=\frac{2 \pi}{m}|X|_{m}
\end{equation}

When the digital operand is in RNS format, satisfying $ X<m $, the OPDC produces the phase value $ \Delta \varphi = 2\pi X / m $, performing the role of a DAC in the phase domain.
Furthermore, when the digital operand $X$ exceeds the modulus $m$, the ODPC performs the digital-to-analog conversion and the format conversion simultaneously.

\begin{figure}[b]
	\centering\includegraphics[width=0.85 \linewidth]{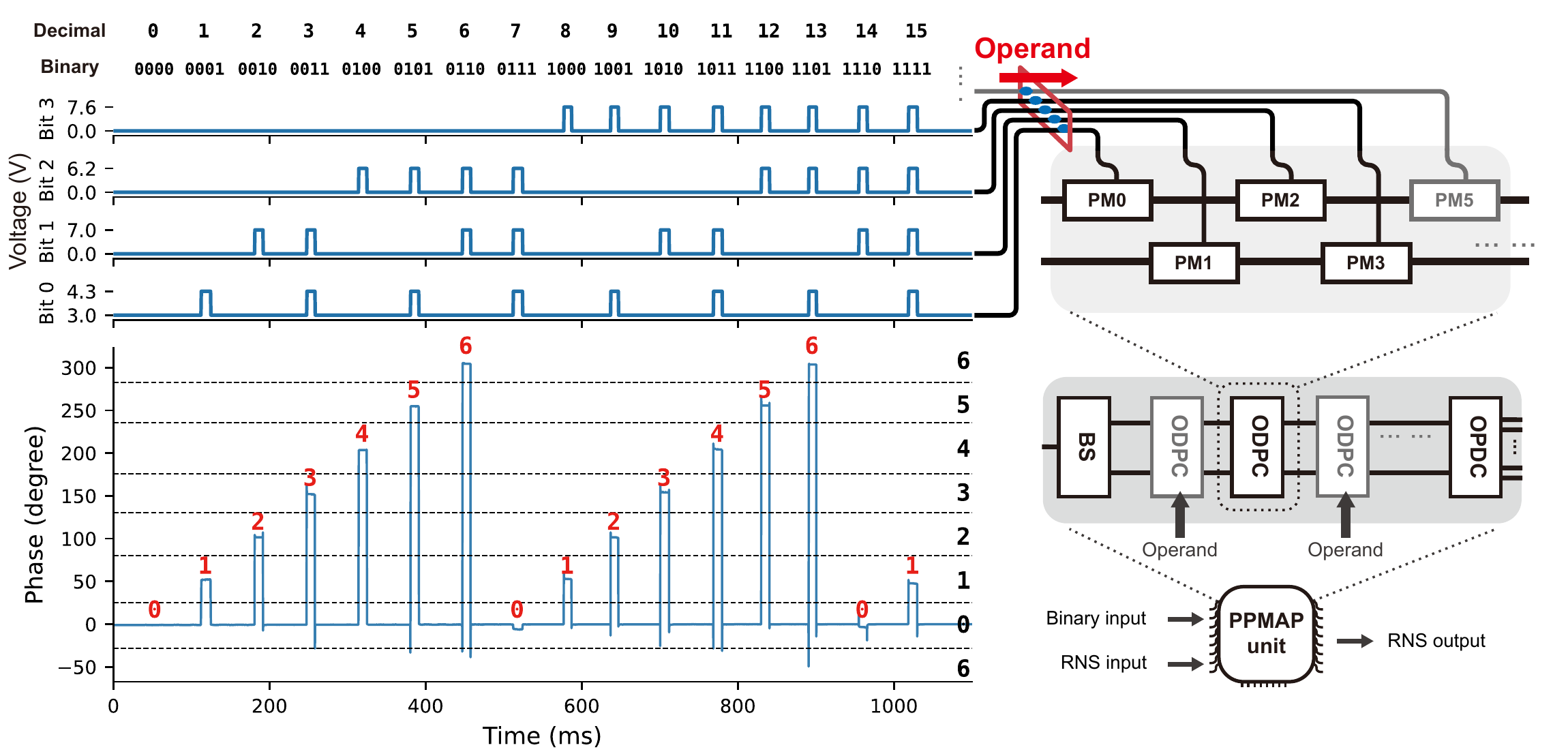}
	\caption{
	Feasibility verification of the photonic conversion interface.
	This experiment demonstrates a PPMAP unit with $m$=7 consisting of one OPDC and one 4-bit ODPC.
	Four PMs are controlled by four parallel digital signals and convert the 4-bit operands into phases.
	In this proof-of-concept experiment, the digital signals are scaled corresponding to the bit position to produce the expected phase change. The digital output of OPDC is also shown in Fig.\ref{figs2}c
	} \label{fig4}
\end{figure}

\begin{figure}[t]
	\centering\includegraphics[width=0.6 \linewidth]{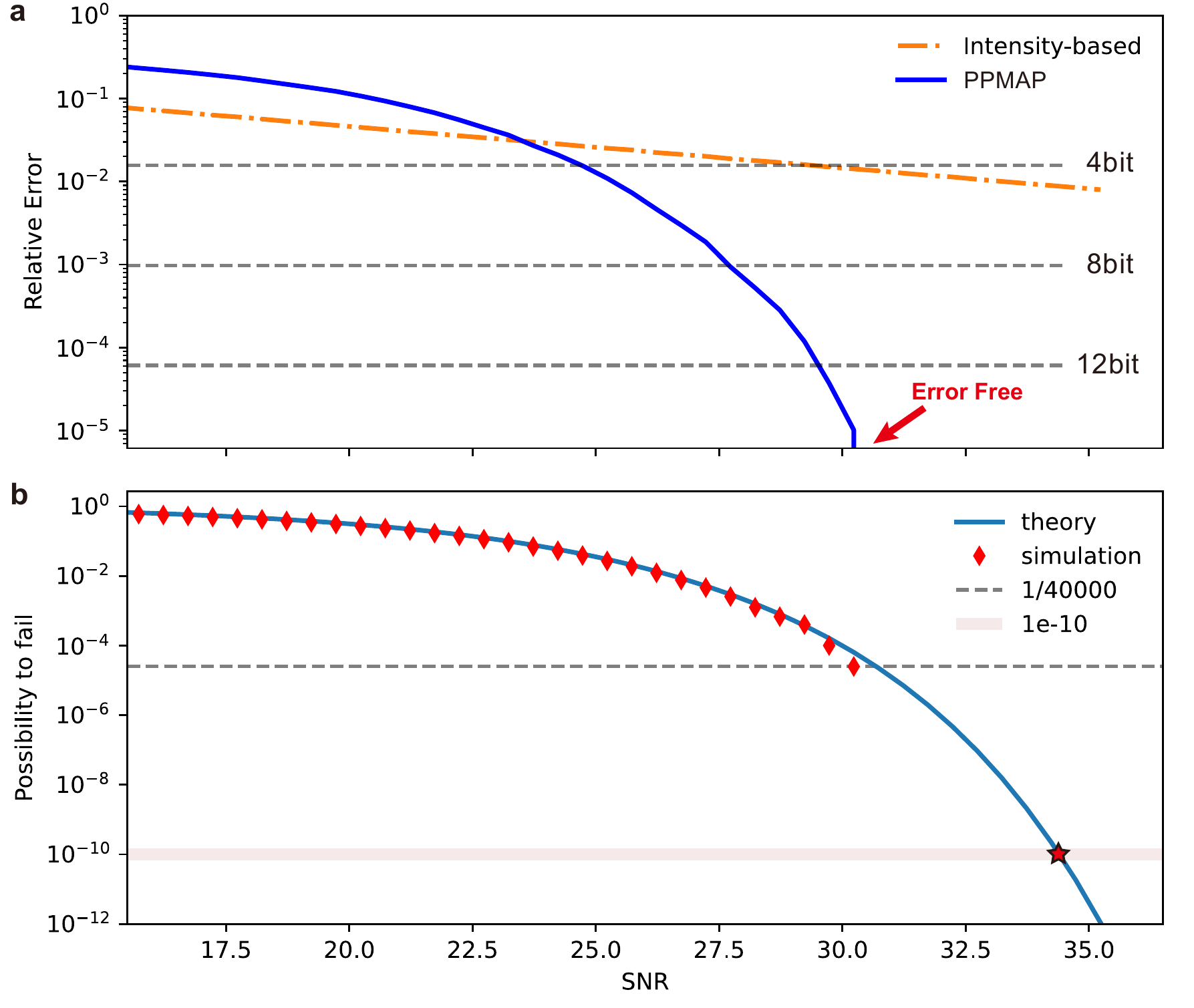}
	\caption
	{
	(a) The relative calculation error of PPMAP and intensity-based optical computing methods in simulating of a 15-bit multiplication task under varying SNR levels. 
	(b) The relative frequency of errors obtained from simulation results for PPMAP, along with its probability model.
	} \label{fig5}
\end{figure}

The experimental demonstration of a 4-bit ODPC along with an OPDC, as shown in Fig.\ref{fig4}, validated the successful operation of this photonic conversion interface.
In this experiment, digital operands representing 4-bit binary codes ranging from 0 to 15 were sequentially transmitted to the corresponding phase modulators PM0, PM1, PM2, and PM3.
With $m=7$, the phase modulators PM1, PM2, and PM3 generated phase changes of 0 or $2\pi |2^i|_7/7$ based on the values of the digital signals (0 or 1) at each bit position $i$, while PM0 generated phase changes of $\varphi_b$ or $\varphi_b+2\pi |2^i|_7/7$, providing an extra bias phase simultaneously.
The experimental results show that the OPDC successfully performed the modulo-7 remainder calculation on the 4-bit binary operands and accurately transferred the results to analog phase values. 

In the generalized ODPC design, low-bit-width DACs are utilized to achieve a larger overall bit width with the given number of PMs (see Appendix \ref{secA3}).
As integrated intensity modulators, such as microring resonators, exhibit poor linearity, achieving accurate mapping of operands with a given precision into transmissivities typically requires the utilization of higher precision DACs \cite{Zhang:22}.
In contrast, the favorable linearity of phase modulation and the characteristics of phase summation enable the decomposition of high-bit-width digital-to-analog conversion tasks into parallel low-bit-width tasks, which can significantly reduce system costs and delays \cite{morales2022analysis} and endow the PPMAP for more efficient processing capabilities.

\section{Discussion}\label{sec3}
The PPMAP architecture offers superior calculation precision compared to intensity-based photonics computing methods. While intensity-based methods heavily rely on the system's SNR for calculation precision, PPMAP leverages the RNS theory to distribute the calculation precision across multiple physical implementations. This decomposition substantially reduces the dependence on SNR and enables a relatively high calculation precision.
Section \ref{subsec1} has demonstrated the superior calculation precision of PPMAP (9 bits) compared to intensity-based methods. 
Now, the focus is on further analyzing the impact of noise on both approaches by conducting simulations to compare their performance in a given 15-bit multiplication scenario.

In the simulations, the specific-setted PPMAP contains four units corresponding to the chosen moduli set of $ \{7, 11, 19, 23\} $ to implement the 15-bit index-sum multiplication.
Each PPMAP unit is equipped with OPDC with $ \{3, 5, 9, 11\} $ channels where additive Gaussian white noise applies to its optical power signal.
Under different SNR conditions, 40,000 pairs of randomly generated integers were selected as the operands for the multiplication operation.
The simulation results presented in Fig.\ref{fig5}a indicate that the relative error and relative frequency of errors in PPMAP decrease significantly with increasing SNR. At an SNR of 31 dB, error-free 15-bit calculation was achieved in all 40,000 samples. On the other hand, the intensity-based method's improvement in relative error with respect to SNR is less prominent, and it falls far short of achieving an 8-bit calculation precision.

Furthermore, to quantify the reliability of the PPMAP architecture, a probability model (Equation \ref{equation_6}) was developed (detailed in Appendix\ref{secA4}):

\begin{equation} \label{equation_6}
    {\rm P} = 1 - \prod_{i=1}^{N} \prod_{j=1}^{K_{i}} \frac{1}{2} \left[1 + {\rm erf}(\frac{\sqrt{3} \sin \left[\frac{\pi}{K_i}(j - \frac{1}{2} )\right]}{ 2 \sqrt{2} \cdot 10^{-\frac{{\rm SNR}}{20} } } ) \right]
\end{equation}

As shown in Fig.\ref{fig5}b, the predicted probabilities of system errors at different SNR levels aligned well with the simulated relative frequency of errors. With the addition of error correction mechanisms, such as Triple Modular Redundancy (TMR) \cite{hochschild2021cores}, the error probability can be reduced to impressively low levels.
For example, with an SNR of approximately 34 dB and employing TMR, the system can achieve an error probability as low as $3 {\rm P}^2 - 2 {\rm P}^3 \approx \text{3e-20}$ \cite{lyons1962use}, indicating a mean time between failure (MTBF) \cite{lienig2017reliability} of around ten years when continuously operating at a frequency of 100 GHz. This level of reliability is comparable with commercial electronic processors \cite{sari2011scrubbing}, which makes PPMAP a promising candidate for performing high-precision and reliable computation tasks.

\section{Conclusion}\label{sec4}
We have proposed PPMAP, an integrated photonic computing architecture that performs modular arithmetic based on the optical phase and RNS.
The architecture is capable of addition, subtraction, and multiplication and supports multi-operand operations.
PPMAP can achieve relatively high precision, and a 9-bit calculation was experimentally demonstrated.
Further simulation extends its calculation precision to 15 bits and indicates that its reliability can approach the level of commercial processors under the SNR condition of 34 dB when using error correction mechanisms.
Moreover, PPMAP seamlessly integrates our proposed ODPC and OPDC, which holds the potential to overcome the long-standing limitations imposed by electronic AD/DA interfaces in photonic computing, and a proof-of-concept experiment demonstrated the feasibility of this photonic conversion interface.
Further optimization of PPMAP's functional module needs to be explored to enhance the overall system performance.

\backmatter

\bmhead{Supplementary information}



The supplementary is located in the Appendix of this article.

\bmhead{Acknowledgments}



\begin{appendices}

\section{Phase Recovery from Multi-Channel Optical Power Signals}\label{secA1}
To recover the phase difference between two paths, $\varphi \triangleq \Delta \varphi + \varphi_0$, from multi-channel optical power signals $\vec{I}$, it is necessary to have knowledge of the corresponding multi-channel optical power states at different values of $\varphi$ beforehand.
That requires utilizing voltage as an intermediate quantity and 
obtaining two mappings:
\begin{enumerate}
	\item Mapping between the applied voltage to PM and resulting $\vec{I}$
	\item Mapping between the voltage and its produced $\varphi$
\end{enumerate}
An experimental procedure was conducted to access the relationship between $\vec{I}$ and $\varphi$. 
The voltage applied to PM0 was varied from 0V to 10V with a voltage increment of 0.01V. Then the first mapping is obtained by recording the multi-channel optical power values at different voltage levels.
The optical power for the $k$-th channel and produced phase variance at voltage index $i$ are denoted as $I_{k,i}$ and $\Delta \varphi_i$.
Moreover, the establishment of the second mapping is based on the prior knowledge of the photonics device.
Theoretically, the variation of output optical power in $K$-channels optical quantizer follows a cosine function with respect to the phase \cite{liu2021experimental} :

\begin{equation}\label{equation_a1}
	\begin{aligned}
		I_{k,i} &= \frac{P_{\rm chip}}{K} \cdot \left( 1 + \cos \varphi^{k}_{i} \right) = \frac{P_{\rm chip}}{K} \cdot \left[ 1 + \cos (\Delta \varphi_i + \varphi_0^{k} ) \right] ,  \\
		\varphi_0^k &=
		\begin{aligned}
			\left\{
			\begin{aligned}
				&\varphi_0 - \frac{k}{2} \cdot \frac{2\pi}{K} & \text{for even }k\\
				&\varphi_0 + \frac{k-1}{2} \cdot \frac{2\pi}{K} & \text{for odd }k\\
			\end{aligned}
			\right.
		\end{aligned}
	\end{aligned}
\end{equation}

, where $P_{\rm chip}$ is the power of the input laser coupled to the chip. The multiple-channel signals are analyzed individually. Firstly, The signal from the $k$-th channel, denoted as $I_{k}$, is scaled to [-1,1] to conform to the input range of the $\text{arccosine}$ function:

\begin{equation}
	\widehat{I}_{k}  = 2 \times \frac{I_{k}}{\max (I_{k})} - 1
\end{equation}

As the $\text{arccosine}$ function is not monotonically changing with respect to the phase $\Delta \varphi$ 
, the relationship of $\varphi_{i}^{k}$ is estimated based on the absolute value of the variation of the $\text{arccos}$ function:

\begin{equation} \label{equation_a3}
	\varphi_{i}^{k}=\left\{
	\begin{aligned}
	&\varphi_i^{k'} \triangleq {\rm arccos} \ \widehat{I}_{k,i} & i=0 \\
	&\varphi_{i-1}^{k} + \vert\varphi_{i+1}^{k'}-\varphi_i^{k'}\vert & i\neq 0
	\end{aligned}
	\right.
\end{equation}

As that accumulation-based approach introduces errors that accumulate over time, a further process is employed to enhance the precision. This process utilizes the ratio of normalized intensity signals with adjacent voltages, denoted as $R_{k,i}=\widehat{I}_{k,i+1}/\widehat{I}_{k,i}$ :

\begin{equation} \label{equation_a4}
	\varphi_i^{k} = \arctan \left[ \frac{\cos\left(\delta_i^k\right)-R_{k,i}}{\sin\left(\delta_i^k\right)}\right] + N \pi
\end{equation}

where $\delta_i^k = \varphi_{i+1}^k-\varphi_i^k$ is the finite difference of phase with respect to voltage,
which along with the integer value $N$ is determined by the previous estimation obtained through Eq.\ref{equation_a3}. 
This process provides finer estimates within the $\pi$ interval based on the coarse-grained results provided in the previous step.
Given the availability of multiple channels, the final step involves an averaging of the phase relationships obtained from each channel :

\begin{equation}
	\varphi = \varphi_0 + \overline{\Delta \varphi_i} = \varphi_0 + \frac{1}{K} \sum_{k = 1}^{K} \varphi_{i}^{k} - \varphi_{0}^{k}  
\end{equation}

With the obtained mapping between $\vec{I}$ and $\varphi$, the phase at a given moment can be determined from the measured optical power.
Expressly, the current $\varphi$ is specified as the phase value corresponding to the previously recorded optical power with the minimum Euclidean norm compared to the currently measured optical power.
It is worth noting that in Fig.\ref{fig2}f, Fig.\ref{fig3}d, and Fig.\ref{fig4}, 
there are waveform spikes observed in the obtained phase signals at slot boundaries. These spikes result in significant deviations in 1 to 2 sampling points.
These deviations are due to synchronization issues related to the multi-channel voltage sources and multi-channel optical detectors in our experimental setup.

\section{OPDC Threshold Optimization}\label{secA2}

Given the multi-channel optical power signals $\vec{I}(\varphi) = \left[I_1(\varphi), I_2(\varphi), \dots,I_K(\varphi)\right]$, which were obtained from the experimental procedure in Appendix\ref{secA1}, 
and the thresholds of comparators for the $K$ channels $\vec{t} = \left[t_1, \dots t_K\right]$. The intersection of $t_k$ with its corresponding $I_k(\varphi)$ determines two boundary points in the 360-degree phase range. 
The sorted boundaries of all channels is denoted as $\vec{p}_{{\rm expt}} = \vec{p}_{{\rm expt}}(\vec{I}, \vec{t})$, which represents the actual phase boundary points.
And in the ideal scenario, the phase boundaries $\vec{p}_{{\rm ideal}}$ should shifted by half interval $\frac{2\pi}{2m}$ to the discrete phase points $\varphi_0^{'} + \Delta \varphi_{\mathbb{Z}}$, where the total static phase $\varphi_0^{'}$ is adjusted by applied an extra bias phase $\varphi_{b}$. 
Thus the ideal phase boundaries can be described as $p_{ideal}^{i} = \varphi_0 + \varphi_b + i \cdot \frac{2\pi}{m} + \frac{\pi}{m}$.

To accurately extract the phase information and reduce the probability of incorrect quantization operation, an optimization procedure was performed to find the optimal values of $\varphi_b$ and $\vec{t}$ that minimize the discrepancy between $\vec{p}_{{\rm ideal}}(\varphi_b)$ and $\vec{p}_{{\rm expt}}(\vec{I}, \vec{t})$. 
This optimization aims to enhance the robustness of OPDCs by placing the most error-prone phase point in a less vulnerable position.
The optimization problem can be formulated as follows:

\begin{equation}
    \left[\varphi_{b}^{\ast }, \vec{t^{\ast}}\right]_{\vec{I}}
    = \arg \min_{\varphi_b, \vec{t}} \ \max_{i}\left\lvert p_{{\rm ideal}}^i(\varphi_b) - p_{{\rm expt}}^i(\vec{I}, \vec{t}) \right\rvert 
\end{equation}

The optimization objective is to minimize the loss $ \triangleq \lVert \vec{p}_{{\rm ideal}}(\varphi_b) - \vec{p}_{{\rm expt}}(\vec{I}, \vec{t}) \rVert_{\infty}$.
Solving this optimization problem using gradient-based methods can be challenging. Therefore, a heuristic optimization method, the differential evolution (DE) algorithm \cite{ahmad2022differential} is employed.
The DE solver provides multiple trial values for $\vec{t}$ at each iteration. A search algorithm was performed for each trial threshold to determine the minimized loss under a specific search range of $\varphi_b$ with a given precision.
Then those minimum values were passed back to the DE solver for further exploration in the next iteration.

\begin{figure}[htbp]
	\centering\includegraphics[width= \linewidth]{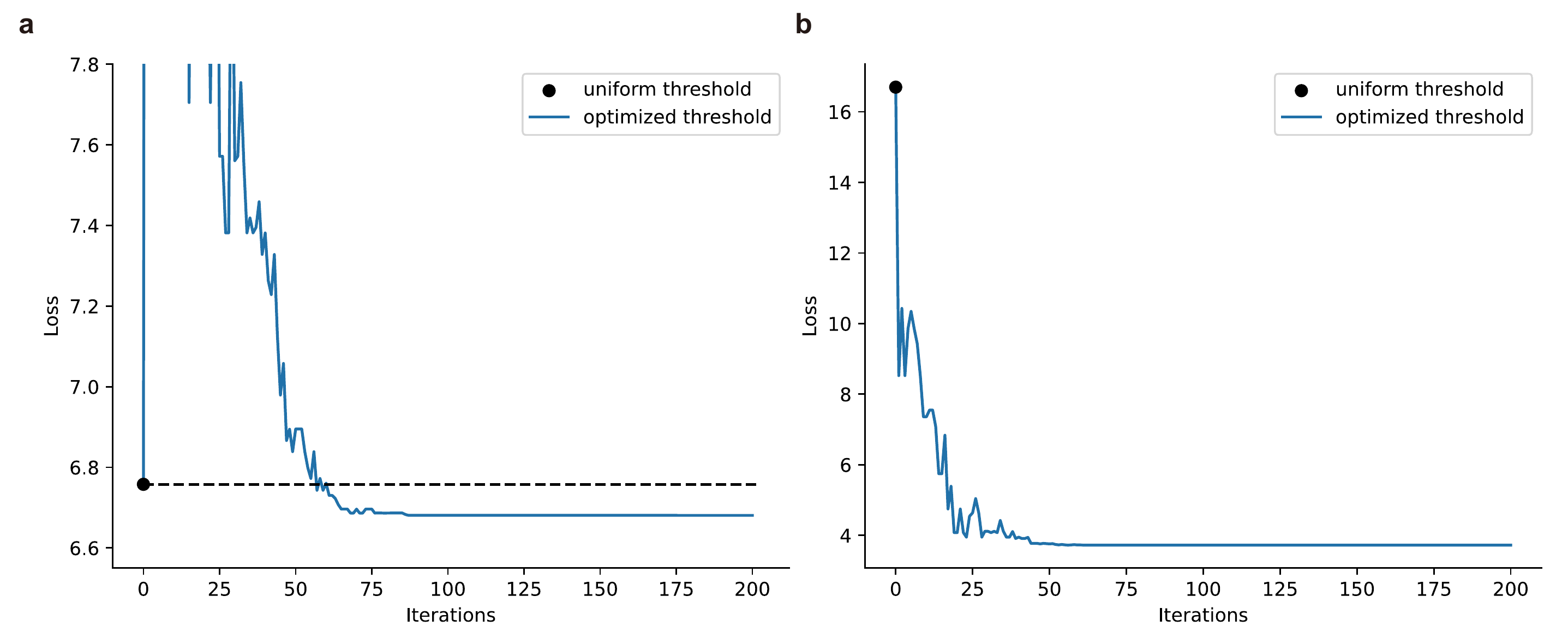}
	\caption
	{
  The process of decision threshold optimization for (a) $m=10$ and (b) $m=7$ through 200 iterations.
  As the iterations progressed, the loss in the optimization objective exhibited a decreasing trend with fluctuations, eventually reaching a relatively stable state.
	} \label{figs1}
\end{figure}

In the practical experiments, the DE solver had a population size of 20 and was run for 40 iterations.
Fig.\ref{figs1} presented the iteration process of threshold optimization for $m=7$ and $m=10$ in the case of 5-channels OPDC.
When modulus $m=2K$, the intervals tend to be evenly divided.
In our previous works \cite{liu2021experimental}, we adopted a uniform threshold approach, where the thresholds for all channels were set to half of the normalized power. This approach equally divides the phase range $2\pi$ into $2K$ intervals in the ideal case. However, as shown in Fig.\ref{figs1}a, in real scenarios where manufacturing variations exist, the optimized threshold approach performs better than the uniform threshold approach.
Moreover, the optimized threshold approach can adopt different moduli with the same optical device.
When modulus $m<2K$ (Fig.\ref{figs1}b), there is a need to merge the remaining phase intervals to unify them into uniformly divided intervals of $m$. This merging process involves mapping multiple different codewords to the same integer.

\begin{figure}[htbp]
	\centering\includegraphics[width=0.8\linewidth]{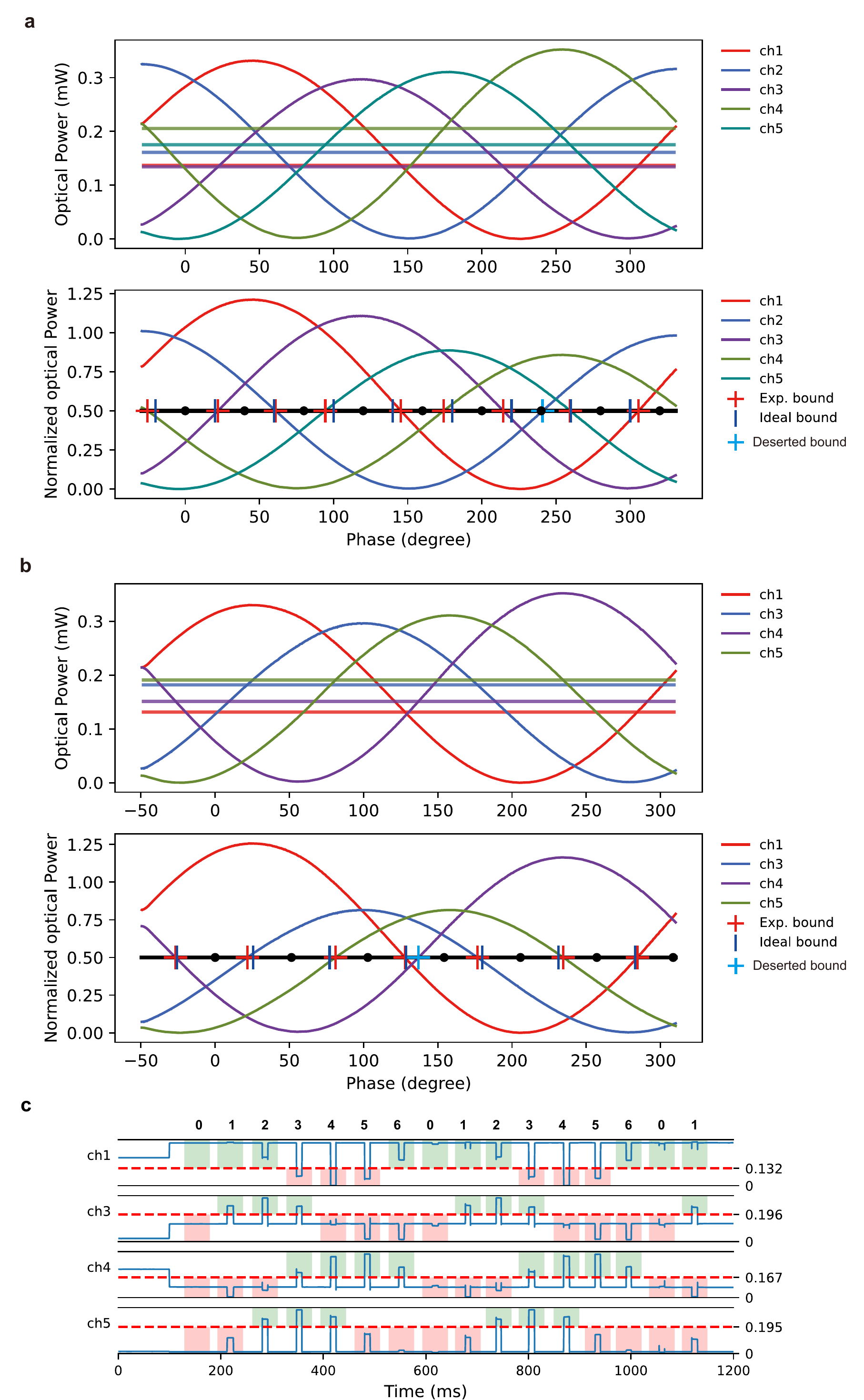}
	\caption
	{
  The relationship between multi-channel optical power values and normalized optical power values with respect to phase change, as well as the optimized thresholds for m=9 (a) and 7 (b).
  The normalization method involved scaling the original optical power values for each channel such that the optimized thresholds were uniformly scaled to 0.5.
  (c) The multi-channel optical power signals and the threshold decision results in the 4-bit ODPC experiment with $m=7$, corresponding to Fig.\ref{fig4}.
	} \label{figs2}
\end{figure}

Fig.\ref{figs2}a,b displays the optimized results for $m=9$ and $m=7$.
It is worth noting that for $m=7$, it is possible to utilize information from 5 optical power signals to quantify the phase information. However, signals from only channels 1, 3, 4, and 5 were employed to reduce the length of the codewords in our experimental setup. The experimental results demonstrate that this approach still yields satisfactory outcomes.
Moreover, in Fig.\ref{figs2}c, we attach the multi-channel optical power signal of the experiment in Section\ref{subsec3}, which uses the optimized parameters corresponding to figure\ref{figs2}b. 
The time-domain waveforms and the digital results of threshold decision confirm the effectiveness of the optimization method.

\section{Generalized Design of ODPC}\label{secA3}
\begin{figure}[htbp]
	\centering\includegraphics[width=\linewidth]{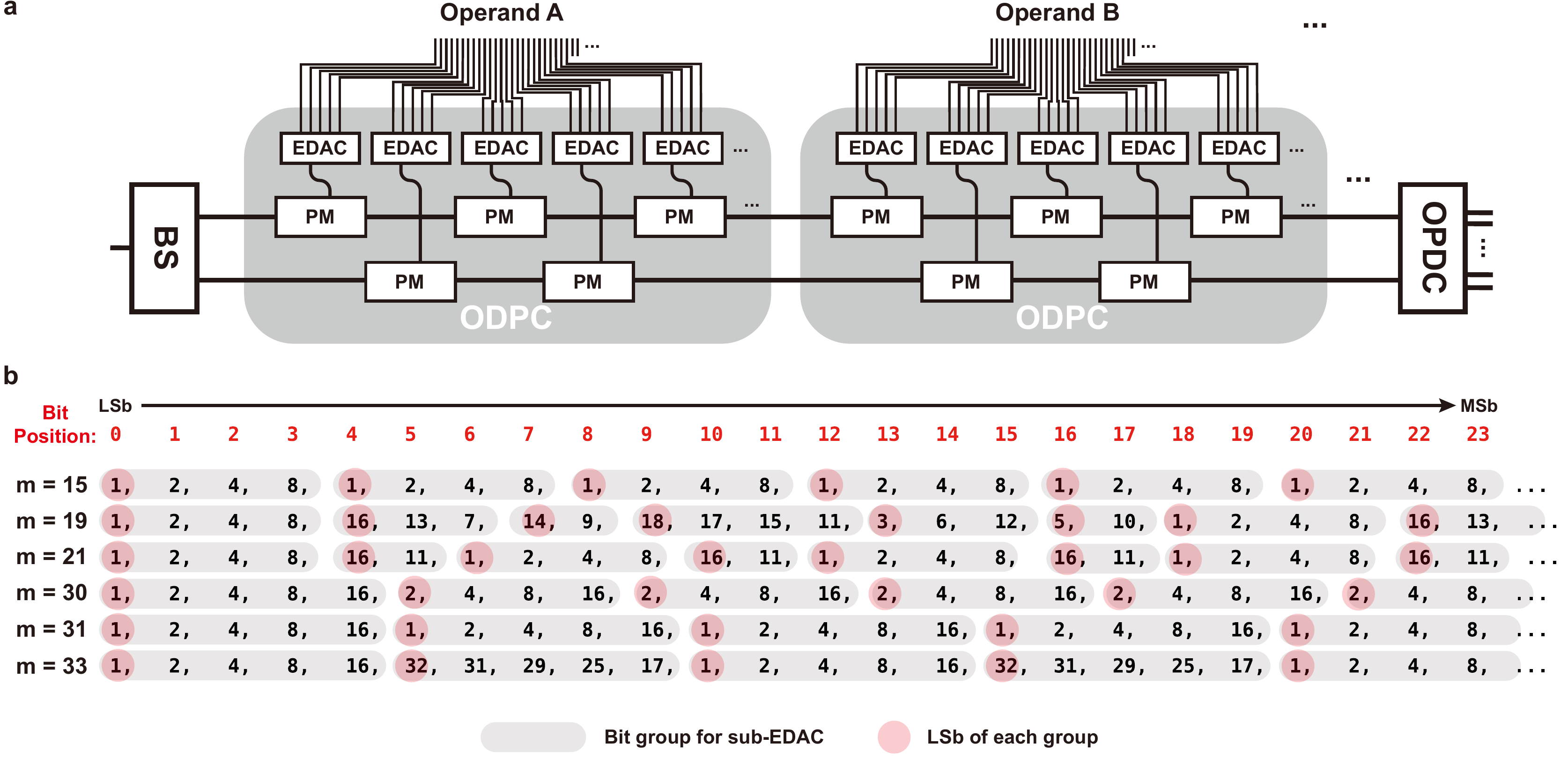}
	\caption
	{
	(a) Structure diagram of a PPMAP unit that adopts the generalized design of OPDC. 
    (b) Example of the grouping strategy of input digital signals, which is based on the bit weight : $\vert 2^{i} \vert_{m}$.
	} \label{figs3}
\end{figure}

The incorporation of high linearity phase modulators, e.g., lithium niobate modulators \cite{Zhu:21}, enhances the capabilities of the ODPC by allowing it to be combined with low-bit-width DACs, thus achieving a larger overall bit width in the limitation of given the number of PMs.
To accomplish this, the $L$-bit digital operand is partitioned into G groups corresponding to G DACs, where the $i$-th DAC has a bit width of $W_i$ (with $\Sigma_{i=1}^{G} W_{i} = L$). The phase response of the $i$-th PM to the least significant bit (LSB) voltage of the $i$-th DAC is set to $ \Delta \varphi_i^{'}=2\pi | 2^{w_{i}} |{m}/m $, where $w_{i}=\Sigma_{j=1}^{i-1} W_{j}$ represents the bit position of the LSB of the i-th DAC in the original L-bit operand.
By mapping the digital operand $X$ using the DACs and the corresponding PMs, the resulting $\Delta \varphi_{X}$ is corresponding in Eq.\ref{equation_3}:

\begin{equation} \label{equation_s6}
	\Delta \varphi_{X}
  =\sum_{i=1}^{G} \sum_{j=0}^{W_{i}-1} 2^j x_{j} \Delta \varphi_i^{'} 
  =\frac{2\pi}{m} \sum_{i=1}^{G} \sum_{j=0}^{W_{i}-1} 2^{w_{i}} | 2^j x_{j} |_{m}
  =\frac{2 \pi}{m} \sum_{j=0}^{L-1} | 2^{j} x_{j}|_{m}
  =\frac{2 \pi}{m}|X|_{m}
\end{equation}
The basic design illustrated in Fig.\ref{fig4} is the specific case in which $W_i$ is limited to 1.
The PPMAP unit, which incorporates the generalized ODPCs, is depicted in Fig.\ref{figs3}a.

For a given electro-optic coefficient of the electro-optic material in the modulator and a given modulus, $m$, the objective of the grouping strategy is to minimize the magnitude of the modulation phase, which contributes to reduceing the size and power consumption of the modulator \cite{Zhu:21}.

According to Eq.\ref{equation_s6}, the LSB phase response of the $i^{{\rm th}}$ modulator for the DAC should be $2\pi | 2^{w_{i}} |_{m}/m$. To achieve this, we need to choose a grouping method for the L-bit digital signal such that $| 2^{w_{i}} |_{m}$ is as close to zero as possible.

Being close to zero implies being in proximity to either 0 or $m$ because a digital value $\vert 2^{i} \vert_{m}$ close to m can be achieved by modulating the phase of the digital value $m-\vert 2^{i} \vert_{m}$ on the opposite waveguide. That is achieved by subtracting a small number from a large number in the modulo operation.
Fig.\ref{figs3}b illustrates examples of bit grouping for moduli of 15, 19, 21, 30, 31, and 33, where the maximum modulation phase of any PM, $(2^{W_i}-1)\times \min \left(\vert 2^{i} \vert_{m}, m-\vert 2^{i} \vert_{m}\right) $, is designed to around $2\pi$. 
We have observed that the bit weights $\vert 2^{i} \vert_{m}$ can be close to zero by employing an appropriate grouping strategy. 
And for some moduli, $| 2^{i} |_{m}$ exhibits periodicity. The specific analysis for the ODAC design depends on the bit width and performance requirements of the DAC, the characteristics of modulators, and the specific modulus.

\section{Analysis of Calculation Precision and Reliability}\label{secA4}
In Fig\ref{fig5}a, the relative error (RE) is defined as the average absolute value to error over the dynamic range.
The RE of a calculation with $n$-bit precision is defined as the $n$-bit relative quantization error of the calculation result: 

\begin{equation}
	{\rm RE}_{n} = \min_{a \in \mathbb{R}} \frac{\int_{a}^{\delta+a} \vert x \vert \cdot \frac{1}{\delta} \,dx  }{\delta \cdot (2^{{n}}-1)} = \frac{1}{2^{{n+2}}-4} \approx 2^{-(n+2)}
\end{equation}

And the theoretical RE of the intensity-based method under Gaussian noise with uniform intensity is given by:

\begin{equation}
	{\rm RE}_{I} = \frac{\int_{-\infty }^{\infty} \vert x \vert f(x) \,dx }{A_{\max}} = \frac{2\int_{0}^{\infty} x\cdot \frac{1}{\sigma \sqrt{2\pi}} e^{\frac{x^2}{2\sigma^{2}}} \,dx }{A_{\max}} = \frac{2}{\sqrt{2\pi}} (\frac{\sigma}{A_{\max}})
\end{equation}

where $f(x)$ represents the probability density function of the normal distribution
and $A_{\max}$ is the maximum signal amplitude. The ratio of $\sigma$ and $A_{\max}$ is determined by its relationship with SNR and peak-to-average power ratio (PAPR):

\begin{equation}
	{\rm SNR} = 20 \log_{10} (\frac{A_{\max}}{\sigma}) - {\rm PAPR}
\end{equation}

Consequently, the relationships of RE between SNR as well as the calculation precision expressed in the form of the bit length, $n$, is inferred:

\begin{equation} \label{equation_d10}
	\log_{10}{\rm RE}_{I} = \log_{10}(\frac{2}{\sqrt{2\pi}}) - \frac{{\rm PAPR}}{20} - \frac{{\rm SNR}}{20} = -(n+2) \cdot \log_{10} 2
\end{equation}

The logarithm of the RE decreases linearly with SNR and $n$. 
Assuming the calculation results follow the uniform distribution, it has a PAPR of 4.77 dB (i.e., $A_{\max} = \sqrt{3} A_{{\rm rms}}$), Eq.\ref{equation_d10} can be simplified to: 

\begin{equation} \label{equation_d11}
	{\rm SNR} \approx 6.02 \cdot n + 5.31
\end{equation}

This equation shows that as $n$ increases, the required SNR increases linearly with a slope of 6.02, corresponding to the numerical simulation shown in Fig\ref{fig5}a. It is worth noting that the constant term in the equation may differ depending on the specific metrics of calculation error used. However, the coefficient of the linear term remains similar.
For example, in the reference \cite{xu202111}, they directly employ ${\rm SNR} = 20 \log_{10} 2^{n} = 6.02 \cdot n$, which aligns with the derived relationship in Eq.\ref{equation_d11}.
That linear term indicates that as the calculation precision increases, the required SNR will reach an unattainable level and imply an exponential increase in energy consumption \cite{tait2022quantifying}.

For the PPMAP architecture, the probability of the occurrence of errors, denoted as ${\rm P}$, is equal to 1 minus that of the PPMAP system successfully operating, which means that there should be no errors in any of the $N$ PPMAP units.
Moreover, the probability of the $i$-th PPMAP unit not encountering errors, denoted as ${\rm P}^{{\rm unit}}_{i}$, is determined by accurately achieving threshold decisions for the output optical power on all the $K_i$ channels. The success rate of the $j$-th channel is represented as ${\rm P}^{\rm chan}_{i,j}$.

\begin{equation} \label{eqd13}
    \begin{aligned}
		{\rm P} = 1 - \prod_{i=1}^{N} {\rm P}_{i}^{{\rm unit}} = 1 - \prod_{i=1}^{N} \prod_{j=1}^{K_{i}} {\rm P}_{i,j}^{{\rm chan}}
	\end{aligned}
\end{equation}

In the simulation, each chosen modulus, i.e., 7, 11, 19, 23, is prime, denoted as $p$. In this case, an index-sum modular multiplication operation is implemented via a modular addition operation with moduli $p-1$ \cite{ramirez2003design}, equal to twice the designed $K_i$ in each PPMAP unit.
That makes for all discrete phase points in $\Delta \varphi_{\mathbb{Z}}$, the relative distance, $\left\lvert d_j \right\rvert / A_{\max}$, between their corresponding ideal output optical power and decision threshold in all $K_i$ channels are given by \cite{tian2018chip}.

\begin{equation}
	\left\{ \frac{\left\lvert d_j \right\rvert}{A_{\max}}  \bigg| \ j\in \mathbb{Z}, 1\leq j \leq K_{i} \right\} = \left\{  \frac{1}{2}\sin \left[ \frac{\pi}{K_{i}}(j^{'} - \frac{1}{2} ) \right] \bigg| \ j^{'} \in \mathbb{Z}, 1\leq j^{'} \leq K_{i}\right\} 
\end{equation}

Considering the relative variance noise $A_{\max}/\sigma = \sqrt{3} A_{{\rm rms}}/\sigma = \sqrt{3} \cdot 10^{\frac{{\rm SNR}}{20}}$,
the success rate for the $j^{'}$-th channel in the $i$-th unit:

\begin{equation}
	{\rm P}_{i,j^{'}}^{{\rm chan}}={\Phi} \left(\frac{\sqrt{3}}{2} \sin \left[ \frac{(j^{'} - \frac{1}{2} ) \pi}{K_i} \right] \cdot 10^{\frac{{\rm SNR}}{20} } \right)
\end{equation}

where ${\Phi}(x) = [1+ ${\rm erf}$(x/\sqrt{2})]/2 $, i.e., the cumulative distribution function of standard Gaussian distribution $\mathcal{N} (0,1)$.
Then considering the commutative law of the continued multiplication in Eq.\ref{eqd13}, $\Pi_{j=1}^{K_{i}} {\rm P}_{i,j}^{{\rm chan}}=\Pi_{j^{'}=1}^{K_{i}} {\rm P}_{i,j^{'}}^{{\rm chan}}$, the error model described in Eq.\ref{equation_6} can be derived.

The deviation between the simulated relative frequency and the probability model around 31 dB in Fig.\ref{fig5} is since only a few samples out of the 40,000 experienced errors near an SNR of 31 dB, which does not satisfy the prerequisites of the law of large numbers.
Apart from this specific range, the relative frequency obtained in our simulations aligns well with our probability model, confirming our model's accuracy. 

\section{The Expansibility of OPDC}\label{secA6}
\begin{figure}[htbp]
	\centering\includegraphics[width=0.8 \linewidth]{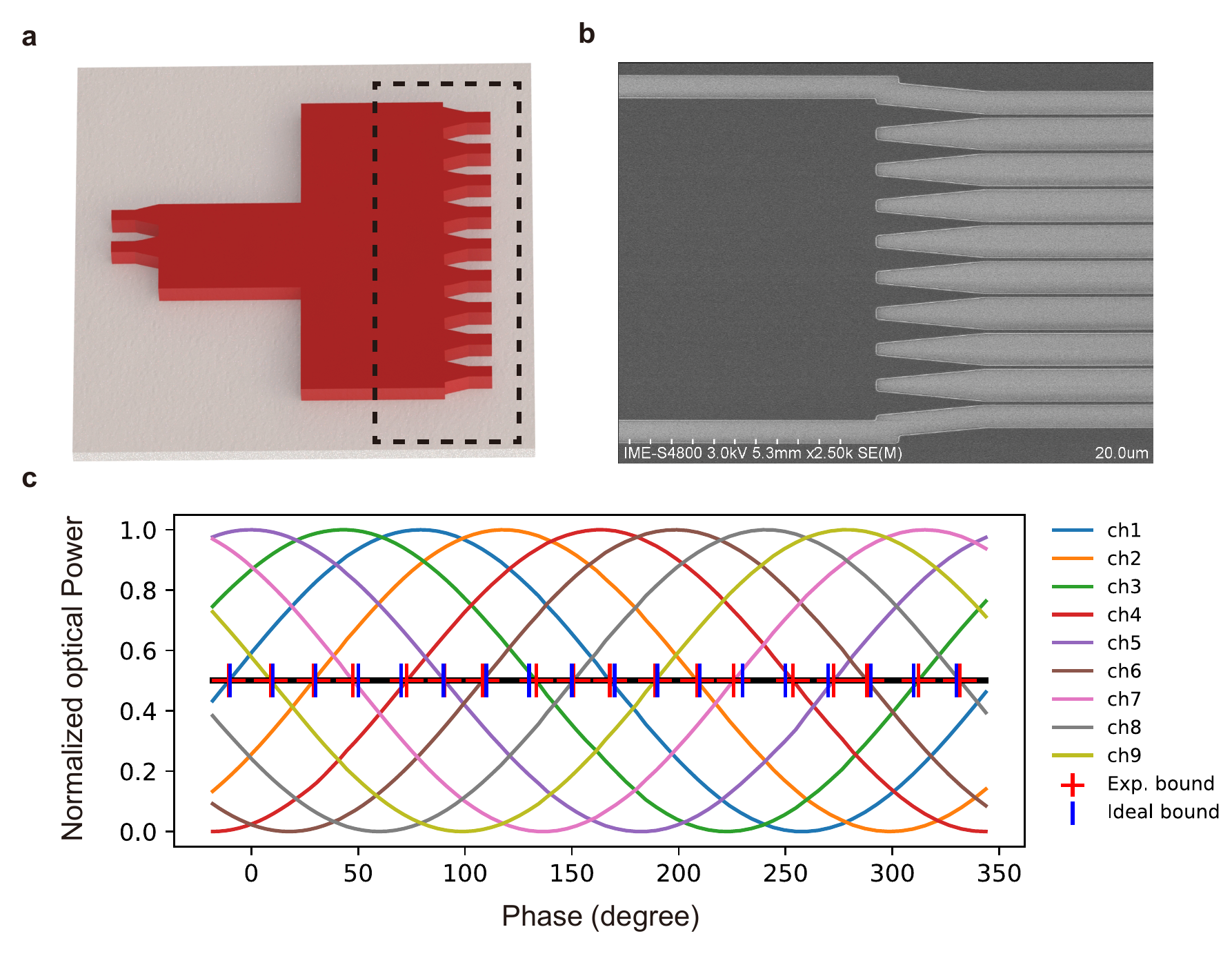}
	\caption
	{
    (a) Structural rendering, (b) micrograph, and (c) the relationship between optical power and phase change of each channel, along with the phase intervals (taking $m=18$ as an example), for the nine-channel OPDC.
	} \label{figs5}
\end{figure}

Although all the above experiment is carried out on the same PPMAP chip with five-channel OPDC, the number of the output channel in OPDC is adjustable. To demonstrate the expansibility of OPDC, we design and fabricate an OPDC module with nine output channels. Fig.\ref{figs5} shows its structural rending, micrograph, and test results. Taking the modulus equal to 18 as an example, the deviation between the experimental and ideal boundaries is relatively small, indicating that the expanded OPDCs can reliably extract the calculation result from phase information.

Besides horizontal expansion by adding the number of output channels, OPDC can be vertically expanded with a cascaded structure \cite{tian2023cascaded}, which possesses relatively low wavelength sensitivity. 
Furthermore, the OPDC fabricated on a thin-film lithium niobate platform exhibits the same performance on phase identifying \cite{tu2023photonic}, indicating the potential for monolithic integration of the overall PPMAP system on lithium niobate process platform.
It should be emphasized that the Gray code generated by OPDC requires an additional codeword conversion unit to transform into the natural binary code 
for subsequent calculations.
However, specific-designed photonic structure \cite{khoram2019nanophotonic, zhu2022space, mohammadi2019inverse, fu2023photonic} has the potential to implement the fixed codeword conversion task. 

\section{Experimental Method}\label{secA5}

We employed a Tunable Semiconductor Laser (Santec TSL 570) to provide the incident CW laser at the wavelength of 1535 nm, and a Multi-port Optical Power Meter (Santec MPM 210) to measure the multi-channel optical power signals.
After the electrical packaging of the integrated photonic chip, metal wires connect the electrodes on the chip to the FPC (Flexible Printed Circuit) adapter board. We employed a multi-channel programmable voltage source (nicslab XDAC-40MUB-R4G8) to control the applied voltages on the electrodes.
The experimental equipment and the host computer were connected via GPIB interface and Ethernet cables, which facilitated the transmission of control signals and the acquisition of experimental data. In-house software was employed to perform real-time monitoring of the optical power signals from each port, which allowed evaluate system stability and monitor the polarization state of the incident light for controlling the coupling efficiency. 


\end{appendices}


\bibliography{sn-bibliography}


\begin{thebibliography}{48}
\ifx \bisbn   \undefined \def \bisbn  #1{ISBN #1}\fi
\ifx \binits  \undefined \def \binits#1{#1}\fi
\ifx \bauthor  \undefined \def \bauthor#1{#1}\fi
\ifx \batitle  \undefined \def \batitle#1{#1}\fi
\ifx \bjtitle  \undefined \def \bjtitle#1{#1}\fi
\ifx \bvolume  \undefined \def \bvolume#1{\textbf{#1}}\fi
\ifx \byear  \undefined \def \byear#1{#1}\fi
\ifx \bissue  \undefined \def \bissue#1{#1}\fi
\ifx \bfpage  \undefined \def \bfpage#1{#1}\fi
\ifx \blpage  \undefined \def \blpage #1{#1}\fi
\ifx \burl  \undefined \def \burl#1{\textsf{#1}}\fi
\ifx \doiurl  \undefined \def \doiurl#1{\url{https://doi.org/#1}}\fi
\ifx \betal  \undefined \def \betal{\textit{et al.}}\fi
\ifx \binstitute  \undefined \def \binstitute#1{#1}\fi
\ifx \binstitutionaled  \undefined \def \binstitutionaled#1{#1}\fi
\ifx \bctitle  \undefined \def \bctitle#1{#1}\fi
\ifx \beditor  \undefined \def \beditor#1{#1}\fi
\ifx \bpublisher  \undefined \def \bpublisher#1{#1}\fi
\ifx \bbtitle  \undefined \def \bbtitle#1{#1}\fi
\ifx \bedition  \undefined \def \bedition#1{#1}\fi
\ifx \bseriesno  \undefined \def \bseriesno#1{#1}\fi
\ifx \blocation  \undefined \def \blocation#1{#1}\fi
\ifx \bsertitle  \undefined \def \bsertitle#1{#1}\fi
\ifx \bsnm \undefined \def \bsnm#1{#1}\fi
\ifx \bsuffix \undefined \def \bsuffix#1{#1}\fi
\ifx \bparticle \undefined \def \bparticle#1{#1}\fi
\ifx \barticle \undefined \def \barticle#1{#1}\fi
\bibcommenthead
\ifx \bconfdate \undefined \def \bconfdate #1{#1}\fi
\ifx \botherref \undefined \def \botherref #1{#1}\fi
\ifx \url \undefined \def \url#1{\textsf{#1}}\fi
\ifx \bchapter \undefined \def \bchapter#1{#1}\fi
\ifx \bbook \undefined \def \bbook#1{#1}\fi
\ifx \bcomment \undefined \def \bcomment#1{#1}\fi
\ifx \oauthor \undefined \def \oauthor#1{#1}\fi
\ifx \citeauthoryear \undefined \def \citeauthoryear#1{#1}\fi
\ifx \endbibitem  \undefined \def \endbibitem {}\fi
\ifx \bconflocation  \undefined \def \bconflocation#1{#1}\fi
\ifx \arxivurl  \undefined \def \arxivurl#1{\textsf{#1}}\fi
\csname PreBibitemsHook\endcsname

\bibitem[\protect\citeauthoryear{Hennessy and
  Patterson}{2018}]{hennessy2018new}
\begin{bchapter}
\bauthor{\bsnm{Hennessy}, \binits{J.}},
\bauthor{\bsnm{Patterson}, \binits{D.}}:
\bctitle{A new golden age for computer architecture: domain-specific
  hardware/software co-design, enhanced}.
In: \bbtitle{ACM/IEEE 45th Annual International Symposium on Computer
  Architecture (ISCA)}
(\byear{2018})
\end{bchapter}
\endbibitem

\bibitem[\protect\citeauthoryear{Esmaeilzadeh et~al.}{2011}]{dark_silicon}
\begin{bchapter}
\bauthor{\bsnm{Esmaeilzadeh}, \binits{H.}},
\bauthor{\bsnm{Blem}, \binits{E.}},
\bauthor{\bsnm{St.~Amant}, \binits{R.}},
\bauthor{\bsnm{Sankaralingam}, \binits{K.}},
\bauthor{\bsnm{Burger}, \binits{D.}}:
\bctitle{Dark silicon and the end of multicore scaling}.
In: \bbtitle{Proceedings of the 38th Annual International Symposium on Computer
  Architecture},
pp. \bfpage{365}--\blpage{376}
(\byear{2011})
\end{bchapter}
\endbibitem

\bibitem[\protect\citeauthoryear{Thompson
  et~al.}{2022}]{thompson2022importance}
\begin{botherref}
\oauthor{\bsnm{Thompson}, \binits{N.C.}},
\oauthor{\bsnm{Ge}, \binits{S.}},
\oauthor{\bsnm{Manso}, \binits{G.F.}}:
The importance of (exponentially more) computing power.
arXiv preprint arXiv:2206.14007
(2022)
\end{botherref}
\endbibitem

\bibitem[\protect\citeauthoryear{Capmany and Novak}{2007}]{2007microwave}
\begin{barticle}
\bauthor{\bsnm{Capmany}, \binits{J.}},
\bauthor{\bsnm{Novak}, \binits{D.}}:
\batitle{Microwave photonics combines two worlds}.
\bjtitle{Nature photonics}
\bvolume{1}(\bissue{6}),
\bfpage{319}
(\byear{2007})
\end{barticle}
\endbibitem

\bibitem[\protect\citeauthoryear{Yang et~al.}{2018}]{yang2018nanowire}
\begin{barticle}
\bauthor{\bsnm{Yang}, \binits{H.}},
\bauthor{\bsnm{Khayrudinov}, \binits{V.}},
\bauthor{\bsnm{Dhaka}, \binits{V.}},
\bauthor{\bsnm{Jiang}, \binits{H.}},
\bauthor{\bsnm{Autere}, \binits{A.}},
\bauthor{\bsnm{Lipsanen}, \binits{H.}},
\bauthor{\bsnm{Sun}, \binits{Z.}},
\bauthor{\bsnm{Jussila}, \binits{H.}}:
\batitle{Nanowire network--based multifunctional all-optical logic gates}.
\bjtitle{Science advances}
\bvolume{4}(\bissue{7}),
\bfpage{7954}
(\byear{2018})
\end{barticle}
\endbibitem

\bibitem[\protect\citeauthoryear{Caballero
  et~al.}{2022}]{caballero2022photonic}
\begin{barticle}
\bauthor{\bsnm{Caballero}, \binits{L.P.}},
\bauthor{\bsnm{Povinelli}, \binits{M.L.}},
\bauthor{\bsnm{Ramirez}, \binits{J.C.}},
\bauthor{\bsnm{Guimar{\~a}es}, \binits{P.S.}},
\bauthor{\bsnm{Neto}, \binits{O.P.V.}}:
\batitle{Photonic crystal integrated logic gates and circuits}.
\bjtitle{Optics Express}
\bvolume{30}(\bissue{2}),
\bfpage{1976}--\blpage{1993}
(\byear{2022})
\end{barticle}
\endbibitem

\bibitem[\protect\citeauthoryear{Ying et~al.}{2020}]{ying2020electronic}
\begin{barticle}
\bauthor{\bsnm{Ying}, \binits{Z.}},
\bauthor{\bsnm{Feng}, \binits{C.}},
\bauthor{\bsnm{Zhao}, \binits{Z.}},
\bauthor{\bsnm{Dhar}, \binits{S.}},
\bauthor{\bsnm{Dalir}, \binits{H.}},
\bauthor{\bsnm{Gu}, \binits{J.}},
\bauthor{\bsnm{Cheng}, \binits{Y.}},
\bauthor{\bsnm{Soref}, \binits{R.}},
\bauthor{\bsnm{Pan}, \binits{D.Z.}},
\bauthor{\bsnm{Chen}, \binits{R.T.}}:
\batitle{Electronic-photonic arithmetic logic unit for high-speed computing}.
\bjtitle{Nature communications}
\bvolume{11}(\bissue{1}),
\bfpage{1}--\blpage{9}
(\byear{2020})
\end{barticle}
\endbibitem

\bibitem[\protect\citeauthoryear{Sebastian et~al.}{2020}]{sebastian2020memory}
\begin{barticle}
\bauthor{\bsnm{Sebastian}, \binits{A.}},
\bauthor{\bsnm{Le~Gallo}, \binits{M.}},
\bauthor{\bsnm{Khaddam-Aljameh}, \binits{R.}},
\bauthor{\bsnm{Eleftheriou}, \binits{E.}}:
\batitle{Memory devices and applications for in-memory computing}.
\bjtitle{Nature nanotechnology}
\bvolume{15}(\bissue{7}),
\bfpage{529}--\blpage{544}
(\byear{2020})
\end{barticle}
\endbibitem

\bibitem[\protect\citeauthoryear{Shafiee et~al.}{2016}]{shafiee2016isaac}
\begin{bchapter}
\bauthor{\bsnm{Shafiee}, \binits{A.}},
\bauthor{\bsnm{Nag}, \binits{A.}},
\bauthor{\bsnm{Muralimanohar}, \binits{N.}},
\bauthor{\bsnm{Balasubramonian}, \binits{R.}},
\bauthor{\bsnm{Strachan}, \binits{J.P.}},
\bauthor{\bsnm{Hu}, \binits{M.}},
\bauthor{\bsnm{Williams}, \binits{R.S.}},
\bauthor{\bsnm{Srikumar}, \binits{V.}}:
\bctitle{Isaac: A convolutional neural network accelerator with in-situ analog
  arithmetic in crossbars}.
In: \bbtitle{2016 ACM/IEEE 43rd Annual International Symposium on Computer
  Architecture (ISCA)},
pp. \bfpage{14}--\blpage{26}
(\byear{2016})
\end{bchapter}
\endbibitem

\bibitem[\protect\citeauthoryear{Mehonic and Kenyon}{2022}]{mehonic2022brain}
\begin{barticle}
\bauthor{\bsnm{Mehonic}, \binits{A.}},
\bauthor{\bsnm{Kenyon}, \binits{A.J.}}:
\batitle{Brain-inspired computing needs a master plan}.
\bjtitle{Nature}
\bvolume{604}(\bissue{7905}),
\bfpage{255}--\blpage{260}
(\byear{2022})
\end{barticle}
\endbibitem

\bibitem[\protect\citeauthoryear{Nahmias et~al.}{2019}]{photonicMAC}
\begin{barticle}
\bauthor{\bsnm{Nahmias}, \binits{M.A.}},
\bauthor{\bsnm{De~Lima}, \binits{T.F.}},
\bauthor{\bsnm{Tait}, \binits{A.N.}},
\bauthor{\bsnm{Peng}, \binits{H.-T.}},
\bauthor{\bsnm{Shastri}, \binits{B.J.}},
\bauthor{\bsnm{Prucnal}, \binits{P.R.}}:
\batitle{Photonic multiply-accumulate operations for neural networks}.
\bjtitle{IEEE Journal of Selected Topics in Quantum Electronics}
\bvolume{26}(\bissue{1}),
\bfpage{1}--\blpage{18}
(\byear{2019})
\end{barticle}
\endbibitem

\bibitem[\protect\citeauthoryear{Filipovich
  et~al.}{2022}]{filipovich2022silicon}
\begin{barticle}
\bauthor{\bsnm{Filipovich}, \binits{M.J.}},
\bauthor{\bsnm{Guo}, \binits{Z.}},
\bauthor{\bsnm{Al-Qadasi}, \binits{M.}},
\bauthor{\bsnm{Marquez}, \binits{B.A.}},
\bauthor{\bsnm{Morison}, \binits{H.D.}},
\bauthor{\bsnm{Sorger}, \binits{V.J.}},
\bauthor{\bsnm{Prucnal}, \binits{P.R.}},
\bauthor{\bsnm{Shekhar}, \binits{S.}},
\bauthor{\bsnm{Shastri}, \binits{B.J.}}:
\batitle{Silicon photonic architecture for training deep neural networks with
  direct feedback alignment}.
\bjtitle{Optica}
\bvolume{9}(\bissue{12}),
\bfpage{1323}--\blpage{1332}
(\byear{2022})
\end{barticle}
\endbibitem

\bibitem[\protect\citeauthoryear{Pai et~al.}{2023}]{pai2023experimental}
\begin{barticle}
\bauthor{\bsnm{Pai}, \binits{S.}},
\bauthor{\bsnm{Park}, \binits{T.}},
\bauthor{\bsnm{Ball}, \binits{M.}},
\bauthor{\bsnm{Penkovsky}, \binits{B.}},
\bauthor{\bsnm{Dubrovsky}, \binits{M.}},
\bauthor{\bsnm{Abebe}, \binits{N.}},
\bauthor{\bsnm{Milanizadeh}, \binits{M.}},
\bauthor{\bsnm{Morichetti}, \binits{F.}},
\bauthor{\bsnm{Melloni}, \binits{A.}},
\bauthor{\bsnm{Fan}, \binits{S.}}, \betal:
\batitle{Experimental evaluation of digitally verifiable photonic computing for
  blockchain and cryptocurrency}.
\bjtitle{Optica}
\bvolume{10}(\bissue{5}),
\bfpage{552}--\blpage{560}
(\byear{2023})
\end{barticle}
\endbibitem

\bibitem[\protect\citeauthoryear{Meng et~al.}{2023}]{meng2023compact}
\begin{barticle}
\bauthor{\bsnm{Meng}, \binits{X.}},
\bauthor{\bsnm{Zhang}, \binits{G.}},
\bauthor{\bsnm{Shi}, \binits{N.}},
\bauthor{\bsnm{Li}, \binits{G.}},
\bauthor{\bsnm{Aza{\~n}a}, \binits{J.}},
\bauthor{\bsnm{Capmany}, \binits{J.}},
\bauthor{\bsnm{Yao}, \binits{J.}},
\bauthor{\bsnm{Shen}, \binits{Y.}},
\bauthor{\bsnm{Li}, \binits{W.}},
\bauthor{\bsnm{Zhu}, \binits{N.}}, \betal:
\batitle{Compact optical convolution processing unit based on multimode
  interference}.
\bjtitle{Nature Communications}
\bvolume{14}(\bissue{1}),
\bfpage{3000}
(\byear{2023})
\end{barticle}
\endbibitem

\bibitem[\protect\citeauthoryear{Zhou et~al.}{2023}]{zhou2023memory}
\begin{barticle}
\bauthor{\bsnm{Zhou}, \binits{W.}},
\bauthor{\bsnm{Dong}, \binits{B.}},
\bauthor{\bsnm{Farmakidis}, \binits{N.}},
\bauthor{\bsnm{Li}, \binits{X.}},
\bauthor{\bsnm{Youngblood}, \binits{N.}},
\bauthor{\bsnm{Huang}, \binits{K.}},
\bauthor{\bsnm{He}, \binits{Y.}},
\bauthor{\bsnm{David~Wright}, \binits{C.}},
\bauthor{\bsnm{Pernice}, \binits{W.H.}},
\bauthor{\bsnm{Bhaskaran}, \binits{H.}}:
\batitle{In-memory photonic dot-product engine with electrically programmable
  weight banks}.
\bjtitle{Nature Communications}
\bvolume{14}(\bissue{1}),
\bfpage{2887}
(\byear{2023})
\end{barticle}
\endbibitem

\bibitem[\protect\citeauthoryear{Xu et~al.}{2022}]{xu2022high}
\begin{barticle}
\bauthor{\bsnm{Xu}, \binits{S.}},
\bauthor{\bsnm{Wang}, \binits{J.}},
\bauthor{\bsnm{Yi}, \binits{S.}},
\bauthor{\bsnm{Zou}, \binits{W.}}:
\batitle{High-order tensor flow processing using integrated photonic circuits}.
\bjtitle{Nature Communications}
\bvolume{13}(\bissue{1}),
\bfpage{7970}
(\byear{2022})
\end{barticle}
\endbibitem

\bibitem[\protect\citeauthoryear{Ashtiani et~al.}{2022}]{nature2022onn}
\begin{botherref}
\oauthor{\bsnm{Ashtiani}, \binits{F.}},
\oauthor{\bsnm{Geers}, \binits{A.J.}},
\oauthor{\bsnm{Aflatouni}, \binits{F.}}:
An on-chip photonic deep neural network for image classification.
Nature,
1--6
(2022)
\end{botherref}
\endbibitem

\bibitem[\protect\citeauthoryear{Shen et~al.}{2017}]{shen2017deep}
\begin{barticle}
\bauthor{\bsnm{Shen}, \binits{Y.}},
\bauthor{\bsnm{Harris}, \binits{N.C.}},
\bauthor{\bsnm{Skirlo}, \binits{S.}},
\bauthor{\bsnm{Prabhu}, \binits{M.}},
\bauthor{\bsnm{Baehr-Jones}, \binits{T.}},
\bauthor{\bsnm{Hochberg}, \binits{M.}},
\bauthor{\bsnm{Sun}, \binits{X.}},
\bauthor{\bsnm{Zhao}, \binits{S.}},
\bauthor{\bsnm{Larochelle}, \binits{H.}},
\bauthor{\bsnm{Englund}, \binits{D.}}, \betal:
\batitle{Deep learning with coherent nanophotonic circuits}.
\bjtitle{Nature photonics}
\bvolume{11}(\bissue{7}),
\bfpage{441}--\blpage{446}
(\byear{2017})
\end{barticle}
\endbibitem

\bibitem[\protect\citeauthoryear{Ashtiani et~al.}{2022}]{ashtiani2022chip}
\begin{barticle}
\bauthor{\bsnm{Ashtiani}, \binits{F.}},
\bauthor{\bsnm{Geers}, \binits{A.J.}},
\bauthor{\bsnm{Aflatouni}, \binits{F.}}:
\batitle{An on-chip photonic deep neural network for image classification}.
\bjtitle{Nature}
\bvolume{606}(\bissue{7914}),
\bfpage{501}--\blpage{506}
(\byear{2022})
\end{barticle}
\endbibitem

\bibitem[\protect\citeauthoryear{Fu et~al.}{2023}]{fu2023photonic}
\begin{barticle}
\bauthor{\bsnm{Fu}, \binits{T.}},
\bauthor{\bsnm{Zang}, \binits{Y.}},
\bauthor{\bsnm{Huang}, \binits{Y.}},
\bauthor{\bsnm{Du}, \binits{Z.}},
\bauthor{\bsnm{Huang}, \binits{H.}},
\bauthor{\bsnm{Hu}, \binits{C.}},
\bauthor{\bsnm{Chen}, \binits{M.}},
\bauthor{\bsnm{Yang}, \binits{S.}},
\bauthor{\bsnm{Chen}, \binits{H.}}:
\batitle{Photonic machine learning with on-chip diffractive optics}.
\bjtitle{Nature Communications}
\bvolume{14}(\bissue{1}),
\bfpage{70}
(\byear{2023})
\end{barticle}
\endbibitem

\bibitem[\protect\citeauthoryear{Tait}{2022}]{tait2022quantifying}
\begin{barticle}
\bauthor{\bsnm{Tait}, \binits{A.N.}}:
\batitle{Quantifying power in silicon photonic neural networks}.
\bjtitle{Physical Review Applied}
\bvolume{17}(\bissue{5}),
\bfpage{054029}
(\byear{2022})
\end{barticle}
\endbibitem

\bibitem[\protect\citeauthoryear{Xu et~al.}{2021}]{xu202111}
\begin{barticle}
\bauthor{\bsnm{Xu}, \binits{X.}},
\bauthor{\bsnm{Tan}, \binits{M.}},
\bauthor{\bsnm{Corcoran}, \binits{B.}},
\bauthor{\bsnm{Wu}, \binits{J.}},
\bauthor{\bsnm{Boes}, \binits{A.}},
\bauthor{\bsnm{Nguyen}, \binits{T.G.}},
\bauthor{\bsnm{Chu}, \binits{S.T.}},
\bauthor{\bsnm{Little}, \binits{B.E.}},
\bauthor{\bsnm{Hicks}, \binits{D.G.}},
\bauthor{\bsnm{Morandotti}, \binits{R.}}, \betal:
\batitle{11 tops photonic convolutional accelerator for optical neural
  networks}.
\bjtitle{Nature}
\bvolume{589}(\bissue{7840}),
\bfpage{44}--\blpage{51}
(\byear{2021})
\end{barticle}
\endbibitem

\bibitem[\protect\citeauthoryear{Krishnamoorthi}{2018}]{krishnamoorthi2018quantizing}
\begin{botherref}
\oauthor{\bsnm{Krishnamoorthi}, \binits{R.}}:
Quantizing deep convolutional networks for efficient inference: A whitepaper.
arXiv preprint arXiv:1806.08342
(2018)
\end{botherref}
\endbibitem

\bibitem[\protect\citeauthoryear{Garg et~al.}{2022}]{garg2022dynamic}
\begin{barticle}
\bauthor{\bsnm{Garg}, \binits{S.}},
\bauthor{\bsnm{Lou}, \binits{J.}},
\bauthor{\bsnm{Jain}, \binits{A.}},
\bauthor{\bsnm{Guo}, \binits{Z.}},
\bauthor{\bsnm{Shastri}, \binits{B.J.}},
\bauthor{\bsnm{Nahmias}, \binits{M.}}:
\batitle{Dynamic precision analog computing for neural networks}.
\bjtitle{IEEE Journal of Selected Topics in Quantum Electronics}
\bvolume{29}(\bissue{2: Optical Computing}),
\bfpage{1}--\blpage{12}
(\byear{2022})
\end{barticle}
\endbibitem

\bibitem[\protect\citeauthoryear{Murmann}{}]{adc_survey}
\begin{botherref}
\oauthor{\bsnm{Murmann}, \binits{B.}}:
{ADC Performance Survey 1997-2023}.
[Online]. Available: \url{https://github.com/bmurmann/ADC-survey}
\end{botherref}
\endbibitem

\bibitem[\protect\citeauthoryear{Morales~Chac{\'o}n
  et~al.}{2022}]{morales2022analysis}
\begin{barticle}
\bauthor{\bsnm{Morales~Chac{\'o}n}, \binits{O.}},
\bauthor{\bsnm{Wikner}, \binits{J.J.}},
\bauthor{\bsnm{Svensson}, \binits{C.}},
\bauthor{\bsnm{Siek}, \binits{L.}},
\bauthor{\bsnm{Alvandpour}, \binits{A.}}:
\batitle{Analysis of energy consumption bounds in cmos current-steering
  digital-to-analog converters}.
\bjtitle{Analog Integrated Circuits and Signal Processing}
\bvolume{111}(\bissue{3}),
\bfpage{339}--\blpage{351}
(\byear{2022})
\end{barticle}
\endbibitem

\bibitem[\protect\citeauthoryear{Dong et~al.}{2022}]{dong2022high}
\begin{barticle}
\bauthor{\bsnm{Dong}, \binits{M.}},
\bauthor{\bsnm{Clark}, \binits{G.}},
\bauthor{\bsnm{Leenheer}, \binits{A.J.}},
\bauthor{\bsnm{Zimmermann}, \binits{M.}},
\bauthor{\bsnm{Dominguez}, \binits{D.}},
\bauthor{\bsnm{Menssen}, \binits{A.J.}},
\bauthor{\bsnm{Heim}, \binits{D.}},
\bauthor{\bsnm{Gilbert}, \binits{G.}},
\bauthor{\bsnm{Englund}, \binits{D.}},
\bauthor{\bsnm{Eichenfield}, \binits{M.}}:
\batitle{High-speed programmable photonic circuits in a cryogenically
  compatible, visible--near-infrared 200 mm cmos architecture}.
\bjtitle{Nature Photonics}
\bvolume{16}(\bissue{1}),
\bfpage{59}--\blpage{65}
(\byear{2022})
\end{barticle}
\endbibitem

\bibitem[\protect\citeauthoryear{Kharel et~al.}{2021}]{kharel2021breaking}
\begin{barticle}
\bauthor{\bsnm{Kharel}, \binits{P.}},
\bauthor{\bsnm{Reimer}, \binits{C.}},
\bauthor{\bsnm{Luke}, \binits{K.}},
\bauthor{\bsnm{He}, \binits{L.}},
\bauthor{\bsnm{Zhang}, \binits{M.}}:
\batitle{Breaking voltage--bandwidth limits in integrated lithium niobate
  modulators using micro-structured electrodes}.
\bjtitle{Optica}
\bvolume{8}(\bissue{3}),
\bfpage{357}--\blpage{363}
(\byear{2021})
\end{barticle}
\endbibitem

\bibitem[\protect\citeauthoryear{Nicholson}{2012}]{nicholson2012introduction}
\begin{bbook}
\bauthor{\bsnm{Nicholson}, \binits{W.K.}}:
\bbtitle{Introduction to Abstract Algebra}.
\bpublisher{John Wiley \& Sons}, \blocation{???}
(\byear{2012})
\end{bbook}
\endbibitem

\bibitem[\protect\citeauthoryear{Omondi and Premkumar}{2007}]{rrns2007residue}
\begin{bbook}
\bauthor{\bsnm{Omondi}, \binits{A.R.}},
\bauthor{\bsnm{Premkumar}, \binits{A.B.}}:
\bbtitle{Residue Number Systems: Theory and Implementation}
vol. \bseriesno{2}.
\bpublisher{World Scientific}, \blocation{???}
(\byear{2007})
\end{bbook}
\endbibitem

\bibitem[\protect\citeauthoryear{Liu et~al.}{2021}]{liu2021experimental}
\begin{barticle}
\bauthor{\bsnm{Liu}, \binits{C.}},
\bauthor{\bsnm{Qiu}, \binits{J.}},
\bauthor{\bsnm{Tian}, \binits{Y.}},
\bauthor{\bsnm{Tao}, \binits{R.}},
\bauthor{\bsnm{Liu}, \binits{Y.}},
\bauthor{\bsnm{He}, \binits{Y.}},
\bauthor{\bsnm{Zhang}, \binits{B.}},
\bauthor{\bsnm{Li}, \binits{Y.}},
\bauthor{\bsnm{Wu}, \binits{J.}}:
\batitle{Experimental demonstration of an optical quantizer with enob of 3.31
  bit by using a cascaded step-size mmi}.
\bjtitle{Optics Express}
\bvolume{29}(\bissue{2}),
\bfpage{2555}--\blpage{2563}
(\byear{2021})
\end{barticle}
\endbibitem

\bibitem[\protect\citeauthoryear{Tian et~al.}{2018}]{tian2018chip}
\begin{barticle}
\bauthor{\bsnm{Tian}, \binits{Y.}},
\bauthor{\bsnm{Qiu}, \binits{J.}},
\bauthor{\bsnm{Huang}, \binits{Z.}},
\bauthor{\bsnm{Qiao}, \binits{Y.}},
\bauthor{\bsnm{Dong}, \binits{Z.}},
\bauthor{\bsnm{Wu}, \binits{J.}}:
\batitle{On-chip integratable all-optical quantizer using cascaded step-size
  mmi}.
\bjtitle{Optics Express}
\bvolume{26}(\bissue{3}),
\bfpage{2453}--\blpage{2461}
(\byear{2018})
\end{barticle}
\endbibitem

\bibitem[\protect\citeauthoryear{Deng et~al.}{2021}]{rrns2021scalable}
\begin{barticle}
\bauthor{\bsnm{Deng}, \binits{B.}},
\bauthor{\bsnm{Srikanth}, \binits{S.}},
\bauthor{\bsnm{Jain}, \binits{A.}},
\bauthor{\bsnm{Conte}, \binits{T.M.}},
\bauthor{\bsnm{DeBenedictis}, \binits{E.}},
\bauthor{\bsnm{Cook}, \binits{J.}}:
\batitle{Scalable energy-efficient microarchitectures with computational error
  tolerance via redundant residue number systems}.
\bjtitle{IEEE Transactions on Computers}
\bvolume{71}(\bissue{3}),
\bfpage{613}--\blpage{627}
(\byear{2021})
\end{barticle}
\endbibitem

\bibitem[\protect\citeauthoryear{Soldano and
  Pennings}{1995}]{soldano1995optical}
\begin{barticle}
\bauthor{\bsnm{Soldano}, \binits{L.B.}},
\bauthor{\bsnm{Pennings}, \binits{E.C.}}:
\batitle{Optical multi-mode interference devices based on self-imaging:
  principles and applications}.
\bjtitle{Journal of lightwave technology}
\bvolume{13}(\bissue{4}),
\bfpage{615}--\blpage{627}
(\byear{1995})
\end{barticle}
\endbibitem

\bibitem[\protect\citeauthoryear{Srikanth et~al.}{2018}]{rrns2018memory}
\begin{bchapter}
\bauthor{\bsnm{Srikanth}, \binits{S.}},
\bauthor{\bsnm{Rabbat}, \binits{P.G.}},
\bauthor{\bsnm{Hein}, \binits{E.R.}},
\bauthor{\bsnm{Deng}, \binits{B.}},
\bauthor{\bsnm{Conte}, \binits{T.M.}},
\bauthor{\bsnm{DeBenedictis}, \binits{E.}},
\bauthor{\bsnm{Cook}, \binits{J.}},
\bauthor{\bsnm{Frank}, \binits{M.P.}}:
\bctitle{Memory system design for ultra low power, computationally error
  resilient processor microarchitectures}.
In: \bbtitle{2018 IEEE International Symposium on High Performance Computer
  Architecture (HPCA)},
pp. \bfpage{696}--\blpage{709}
(\byear{2018}).
\bcomment{IEEE}
\end{bchapter}
\endbibitem

\bibitem[\protect\citeauthoryear{Zhang et~al.}{2022}]{Zhang:22}
\begin{barticle}
\bauthor{\bsnm{Zhang}, \binits{W.}},
\bauthor{\bsnm{Huang}, \binits{C.}},
\bauthor{\bsnm{Peng}, \binits{H.-T.}},
\bauthor{\bsnm{Bilodeau}, \binits{S.}},
\bauthor{\bsnm{Jha}, \binits{A.}},
\bauthor{\bsnm{Blow}, \binits{E.}},
\bauthor{\bsnm{Lima}, \binits{T.F.}},
\bauthor{\bsnm{Shastri}, \binits{B.J.}},
\bauthor{\bsnm{Prucnal}, \binits{P.}}:
\batitle{Silicon microring synapses enable photonic deep learning beyond 9-bit
  precision}.
\bjtitle{Optica}
\bvolume{9}(\bissue{5}),
\bfpage{579}--\blpage{584}
(\byear{2022})
\doiurl{10.1364/OPTICA.446100}
\end{barticle}
\endbibitem

\bibitem[\protect\citeauthoryear{Hochschild et~al.}{2021}]{hochschild2021cores}
\begin{bchapter}
\bauthor{\bsnm{Hochschild}, \binits{P.H.}},
\bauthor{\bsnm{Turner}, \binits{P.}},
\bauthor{\bsnm{Mogul}, \binits{J.C.}},
\bauthor{\bsnm{Govindaraju}, \binits{R.}},
\bauthor{\bsnm{Ranganathan}, \binits{P.}},
\bauthor{\bsnm{Culler}, \binits{D.E.}},
\bauthor{\bsnm{Vahdat}, \binits{A.}}:
\bctitle{Cores that don't count}.
In: \bbtitle{Proceedings of the Workshop on Hot Topics in Operating Systems},
pp. \bfpage{9}--\blpage{16}
(\byear{2021})
\end{bchapter}
\endbibitem

\bibitem[\protect\citeauthoryear{Lyons and Vanderkulk}{1962}]{lyons1962use}
\begin{barticle}
\bauthor{\bsnm{Lyons}, \binits{R.E.}},
\bauthor{\bsnm{Vanderkulk}, \binits{W.}}:
\batitle{The use of triple-modular redundancy to improve computer reliability}.
\bjtitle{IBM journal of research and development}
\bvolume{6}(\bissue{2}),
\bfpage{200}--\blpage{209}
(\byear{1962})
\end{barticle}
\endbibitem

\bibitem[\protect\citeauthoryear{Lienig et~al.}{2017}]{lienig2017reliability}
\begin{botherref}
\oauthor{\bsnm{Lienig}, \binits{J.}},
\oauthor{\bsnm{Bruemmer}, \binits{H.}},
\oauthor{\bsnm{Lienig}, \binits{J.}},
\oauthor{\bsnm{Bruemmer}, \binits{H.}}:
Reliability analysis.
Fundamentals of electronic systems design,
45--73
(2017)
\end{botherref}
\endbibitem

\bibitem[\protect\citeauthoryear{Sari and Psarakis}{2011}]{sari2011scrubbing}
\begin{bchapter}
\bauthor{\bsnm{Sari}, \binits{A.}},
\bauthor{\bsnm{Psarakis}, \binits{M.}}:
\bctitle{Scrubbing-based seu mitigation approach for
  systems-on-programmable-chips}.
In: \bbtitle{2011 International Conference on Field-Programmable Technology},
pp. \bfpage{1}--\blpage{8}
(\byear{2011}).
\bcomment{IEEE}
\end{bchapter}
\endbibitem

\bibitem[\protect\citeauthoryear{Khoram et~al.}{2019}]{khoram2019nanophotonic}
\begin{barticle}
\bauthor{\bsnm{Khoram}, \binits{E.}},
\bauthor{\bsnm{Chen}, \binits{A.}},
\bauthor{\bsnm{Liu}, \binits{D.}},
\bauthor{\bsnm{Ying}, \binits{L.}},
\bauthor{\bsnm{Wang}, \binits{Q.}},
\bauthor{\bsnm{Yuan}, \binits{M.}},
\bauthor{\bsnm{Yu}, \binits{Z.}}:
\batitle{Nanophotonic media for artificial neural inference}.
\bjtitle{Photonics Research}
\bvolume{7}(\bissue{8}),
\bfpage{823}--\blpage{827}
(\byear{2019})
\end{barticle}
\endbibitem

\bibitem[\protect\citeauthoryear{Zhu et~al.}{2022}]{zhu2022space}
\begin{barticle}
\bauthor{\bsnm{Zhu}, \binits{H.}},
\bauthor{\bsnm{Zou}, \binits{J.}},
\bauthor{\bsnm{Zhang}, \binits{H.}},
\bauthor{\bsnm{Shi}, \binits{Y.}},
\bauthor{\bsnm{Luo}, \binits{S.}},
\bauthor{\bsnm{Wang}, \binits{N.}},
\bauthor{\bsnm{Cai}, \binits{H.}},
\bauthor{\bsnm{Wan}, \binits{L.}},
\bauthor{\bsnm{Wang}, \binits{B.}},
\bauthor{\bsnm{Jiang}, \binits{X.}}, \betal:
\batitle{Space-efficient optical computing with an integrated chip diffractive
  neural network}.
\bjtitle{Nature communications}
\bvolume{13}(\bissue{1}),
\bfpage{1044}
(\byear{2022})
\end{barticle}
\endbibitem

\bibitem[\protect\citeauthoryear{Mohammadi~Estakhri
  et~al.}{2019}]{mohammadi2019inverse}
\begin{barticle}
\bauthor{\bsnm{Mohammadi~Estakhri}, \binits{N.}},
\bauthor{\bsnm{Edwards}, \binits{B.}},
\bauthor{\bsnm{Engheta}, \binits{N.}}:
\batitle{Inverse-designed metastructures that solve equations}.
\bjtitle{Science}
\bvolume{363}(\bissue{6433}),
\bfpage{1333}--\blpage{1338}
(\byear{2019})
\end{barticle}
\endbibitem

\bibitem[\protect\citeauthoryear{Ahmad et~al.}{2022}]{ahmad2022differential}
\begin{barticle}
\bauthor{\bsnm{Ahmad}, \binits{M.F.}},
\bauthor{\bsnm{Isa}, \binits{N.A.M.}},
\bauthor{\bsnm{Lim}, \binits{W.H.}},
\bauthor{\bsnm{Ang}, \binits{K.M.}}:
\batitle{Differential evolution: A recent review based on state-of-the-art
  works}.
\bjtitle{Alexandria Engineering Journal}
\bvolume{61}(\bissue{5}),
\bfpage{3831}--\blpage{3872}
(\byear{2022})
\end{barticle}
\endbibitem

\bibitem[\protect\citeauthoryear{Zhu et~al.}{2021}]{Zhu:21}
\begin{barticle}
\bauthor{\bsnm{Zhu}, \binits{D.}},
\bauthor{\bsnm{Shao}, \binits{L.}},
\bauthor{\bsnm{Yu}, \binits{M.}},
\bauthor{\bsnm{Cheng}, \binits{R.}},
\bauthor{\bsnm{Desiatov}, \binits{B.}},
\bauthor{\bsnm{Xin}, \binits{C.J.}},
\bauthor{\bsnm{Hu}, \binits{Y.}},
\bauthor{\bsnm{Holzgrafe}, \binits{J.}},
\bauthor{\bsnm{Ghosh}, \binits{S.}},
\bauthor{\bsnm{Shams-Ansari}, \binits{A.}},
\bauthor{\bsnm{Puma}, \binits{E.}},
\bauthor{\bsnm{Sinclair}, \binits{N.}},
\bauthor{\bsnm{Reimer}, \binits{C.}},
\bauthor{\bsnm{Zhang}, \binits{M.}},
\bauthor{\bsnm{Lon\v{c}ar}, \binits{M.}}:
\batitle{Integrated photonics on thin-film lithium niobate}.
\bjtitle{Adv. Opt. Photon.}
\bvolume{13}(\bissue{2}),
\bfpage{242}--\blpage{352}
(\byear{2021})
\doiurl{10.1364/AOP.411024}
\end{barticle}
\endbibitem

\bibitem[\protect\citeauthoryear{Ram{\'\i}rez et~al.}{2003}]{ramirez2003design}
\begin{barticle}
\bauthor{\bsnm{Ram{\'\i}rez}, \binits{J.}},
\bauthor{\bsnm{Meyer-B{\"a}se}, \binits{U.}},
\bauthor{\bsnm{Taylor}, \binits{F.}},
\bauthor{\bsnm{Garc{\'\i}a}, \binits{A.}},
\bauthor{\bsnm{Lloris}, \binits{A.}}:
\batitle{Design and implementation of high-performance rns wavelet processors
  using custom ic technologies}.
\bjtitle{Journal of VLSI signal processing systems for signal, image and video
  technology}
\bvolume{34},
\bfpage{227}--\blpage{237}
(\byear{2003})
\end{barticle}
\endbibitem

\bibitem[\protect\citeauthoryear{Tian et~al.}{2023}]{tian2023cascaded}
\begin{barticle}
\bauthor{\bsnm{Tian}, \binits{Y.}},
\bauthor{\bsnm{Kang}, \binits{Z.}},
\bauthor{\bsnm{He}, \binits{J.}},
\bauthor{\bsnm{Zheng}, \binits{Z.}},
\bauthor{\bsnm{Qiu}, \binits{J.}},
\bauthor{\bsnm{Wu}, \binits{J.}},
\bauthor{\bsnm{Zhang}, \binits{X.}}:
\batitle{Cascaded all-optical quantization employing step-size mmi and
  shape-optimized power splitter}.
\bjtitle{Optics \& Laser Technology}
\bvolume{158},
\bfpage{108820}
(\byear{2023})
\end{barticle}
\endbibitem

\bibitem[\protect\citeauthoryear{Tu et~al.}{2023}]{tu2023photonic}
\begin{barticle}
\bauthor{\bsnm{Tu}, \binits{D.}},
\bauthor{\bsnm{Huang}, \binits{X.}},
\bauthor{\bsnm{Yu}, \binits{H.}},
\bauthor{\bsnm{Yin}, \binits{Y.}},
\bauthor{\bsnm{Yu}, \binits{Z.}},
\bauthor{\bsnm{Wei}, \binits{Z.}},
\bauthor{\bsnm{Li}, \binits{Z.}}:
\batitle{Photonic sampled and quantized analog-to-digital converters on
  thin-film lithium niobate platform}.
\bjtitle{Optics Express}
\bvolume{31}(\bissue{2}),
\bfpage{1931}--\blpage{1942}
(\byear{2023})
\end{barticle}
\endbibitem

\end{thebibliography}

\end{document}